\documentclass[10pt,letterpaper,twocolumn,english,final,floatfix,showpacs,showkeys,preprintnumbers,prd,nofootinbib,superscriptaddress]{revtex4-1}
\usepackage{color}
\usepackage{float}
\usepackage{graphicx}

\makeatletter

\pdfpageheight\paperheight
\pdfpagewidth\paperwidth

 \@ifundefined{textcolor}{}
 {%
   \definecolor{BLACK}{gray}{0}
   \definecolor{WHITE}{gray}{1}
   \definecolor{RED}{rgb}{1,0,0}
   \definecolor{GREEN}{rgb}{0,1,0}
   \definecolor{BLUE}{rgb}{0,0,1}
   \definecolor{CYAN}{cmyk}{1,0,0,0}
   \definecolor{MAGENTA}{cmyk}{0,1,0,0}
   \definecolor{YELLOW}{cmyk}{0,0,1,0}
 }

\@ifundefined{definecolor}
 {\usepackage{color}}{}
\usepackage{babel}

\newcommand{\be}{\begin{equation}}
\newcommand{\ee}{\end{equation}}
\newcommand{\ba}{\begin{eqnarray}}
\newcommand{\ea}{\end{eqnarray}}

\@ifundefined{showcaptionsetup}{}{%
 \PassOptionsToPackage{caption=false}{subfig}}
\usepackage{subfig}
\makeatother

\begin{document} 

\title{\null \vspace{0.5in}
Gluon shadowing effects on $J/\psi$ and $\Upsilon$ production in $p+$Pb 
collisions at $\sqrt{s_{_{NN}}} = 115$ GeV and Pb$+p$ collisions at 
$\sqrt{s_{_{NN}}} = 72$ GeV at AFTER@LHC}

\author{R. Vogt}
\thanks{vogt2@llnl.gov}
\affiliation{
Nuclear and Chemical Sciences Division, Lawrence Livermore National Laboratory, 
Livermore, CA 94551, USA \break
Physics Department, University of California at Davis, Davis, CA 95616, 
USA
}

\keywords{quarkonium, cold nuclear matter effects}


\begin{abstract}
We explore the effects of shadowing on inclusive $J/\psi$ and $\Upsilon(1S)$
production at AFTER@LHC.  We also present the rates as a function of $p_T$
and rapidity for $p+$Pb and Pb$+p$ collisions in the proposed AFTER@LHC 
rapidity acceptance.
\end{abstract}

\maketitle



\section{Introduction}
\label{sec:intro}

The AFTER@LHC quarkonium program has the unique opportunity to study quarkonium 
production at large
momentum fractions, $x$, in the target region \cite{Brodsky:2012vg}.  
The most favorable 
configuration for high rates at large $x$ for the nucleus is a proton beam
from the LHC 
on a heavy nuclear target.  In this case, the nucleon-nucleon
center of mass energy is
more than half that of the RHIC collider, $\sqrt{s_{_{NN}}} = 115$ GeV, for the
top LHC proton beam energy of 7 TeV. 
However, the fixed-target configuration is an advantage because of the
higher intensity on target.  The
longer LHC proton runs gives a luminosity over a $10^7$ s LHC 'year'.
On a 1 cm thick Pb target, with $p+$Pb collisions, 
${\mathcal L} = 16 A$ $\mu$b$^{-1}$ s$^{-1}$.
When a lead beam is extracted the run time is shorter, an LHC Pb 'year'
is $10^6$ s.  The lower $Z/A$ ratio also results in a lower center-of-mass
energy of $\sqrt{s_{_{NN}}} = 72$ GeV for the top lead beam energy of 2.76 TeV.  
On a liquid H$_2$ target, for Pb$+p$ 
collisions, ${\mathcal L} = 8A_{\rm Pb}$ mb$^{-1}$ s$^{-1}$ 
per centimeter target length so that a 1 m target gives a luminosity of 
${\mathcal L} = 800 A_{\rm Pb}$ mb$^{-1}$ s$^{-1}$ \cite{Brodsky:2012vg}.

Here we will consider the inclusive $J/\psi$ and $\Upsilon$(1S) rates in
$p+$Pb collisions at $\sqrt{s_{_{NN}}} = 115$ GeV and Pb$+p$ collisions at
$\sqrt{s_{_{NN}}}= 72$ GeV.  The results are presented as a function of rapidity,
$y$, and transverse momentum, $p_T$, of the quarkonium state.
We choose to present the $p_T$ results in a 0.5 unit wide rapidity bin in the
backward region of the center of mass of the collision,
$-2.5 < y_{\rm cms} < -2.0$ for $\sqrt{s_{_{NN}}} = 115$ GeV and 
$-1.9 < y_{\rm cms} < -1.4$ for $\sqrt{s_{_{NN}}} = 72$ GeV.  This is a region that 
has been virtually unexplored in previous quarkonium production measurements
but, as we will show, can be studied by AFTER@LHC with relatively high 
statistics in most cases.

Our calculations are done in the next-to-leading
order (NLO) color evaporation model (CEM) \cite{Gavai:1994in} and employ
the EPS09 NLO parameterization \cite{Eskola:2009}
of the effects of modification of the parton
distribution functions in the nucleus, referred to here as 'shadowing'.
Since this set also provides an uncertainty band, 
the results are representative of the range of shadowing parameterizations
produced by other groups.

We also present the nuclear suppression factor
ratios, $R_{p{\rm Pb}}$ for $p+$Pb collisions and
$R_{{\rm Pb}p}$ for Pb$+p$ collisions.  These quantities are the ratio of the 
per nucleon cross sections in $p+$Pb (Pb$+p$) collisions relative the same
cross section in $p+p$ collisions at the same center of mass energy.  
These ratios are also given as a function of $p_T$ and $y$.

In Sec.~\ref{nPDFs}, we will show the EPS09 NLO shadowing parameterizations
at the appropriate factorization scale for $J/\psi$ and $\Upsilon$
production as a function of $x$ with emphasis on the appropriate $x$ regions 
for the AFTER@LHC kinematics.  We present the ratios and rates obtained with the
EPS09 NLO parameterization in Sec.~\ref{rates}.  We conclude with some final
remarks in Sec.~\ref{summary}.

\section{Shadowing parameterization}
\label{nPDFs}

Our calculations employ the EPS09 shadowing parameterization 
\cite{Eskola:2009}.  At NLO, it is based on the CTEQ6M proton parton densities
(PDFs) \cite{CTEQ6M}.  In our calculations of quarkonium production
\cite{Nelson:2012bc,Nelson_inprog}, we use the
CT10 \cite{CT10PDFs} proton PDFs with the EPS09 NLO parameterization.  
As long as both calculations are at NLO, the choice of proton PDFs used to 
calculate quarkonium production does not affect the shape or magnitude of the
nuclear suppression factors \cite{RV_inprog}.

\begin{figure*}[thp!]
\begin{center}
\includegraphics[width=0.495\textwidth]{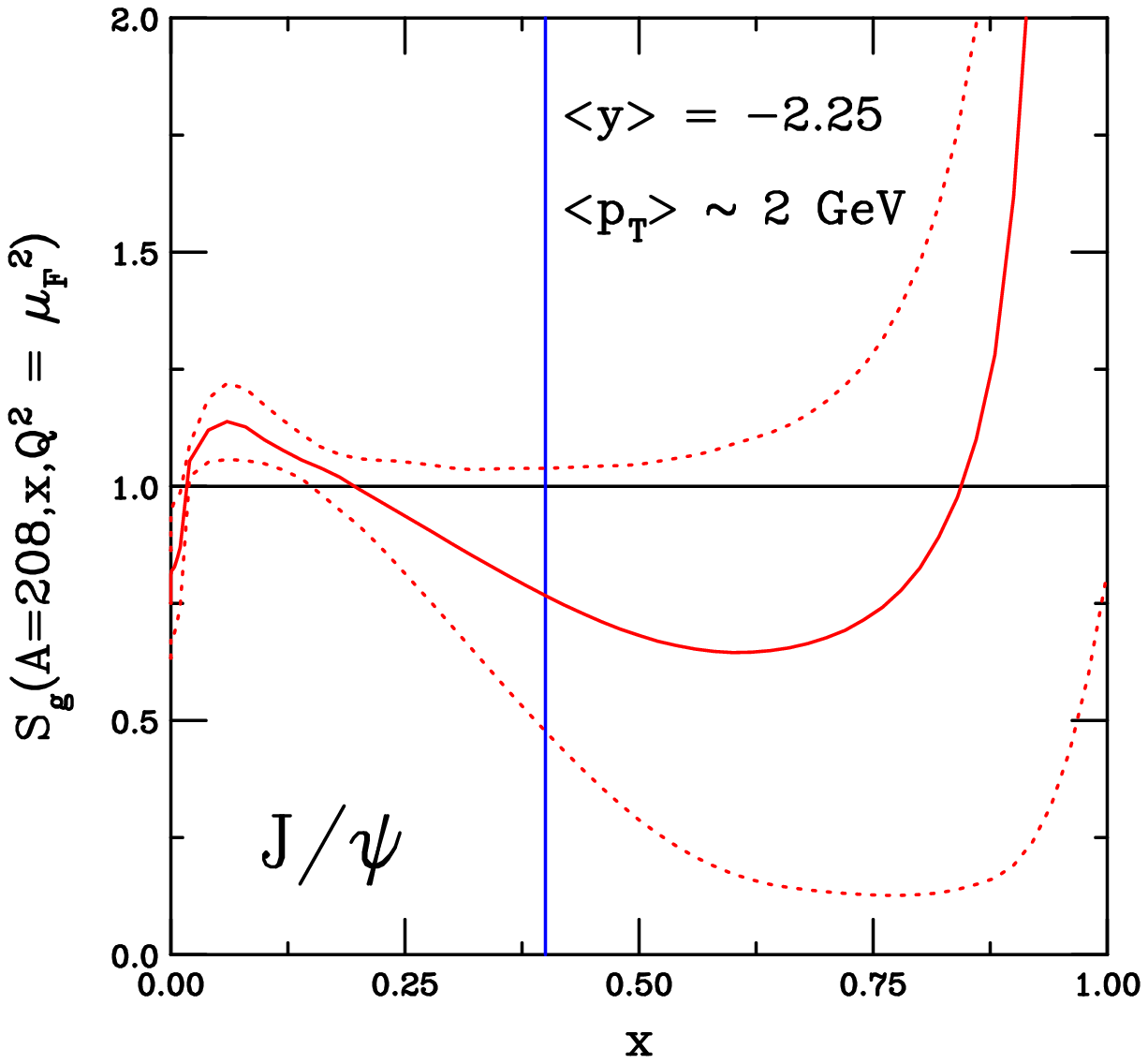}
\includegraphics[width=0.495\textwidth]{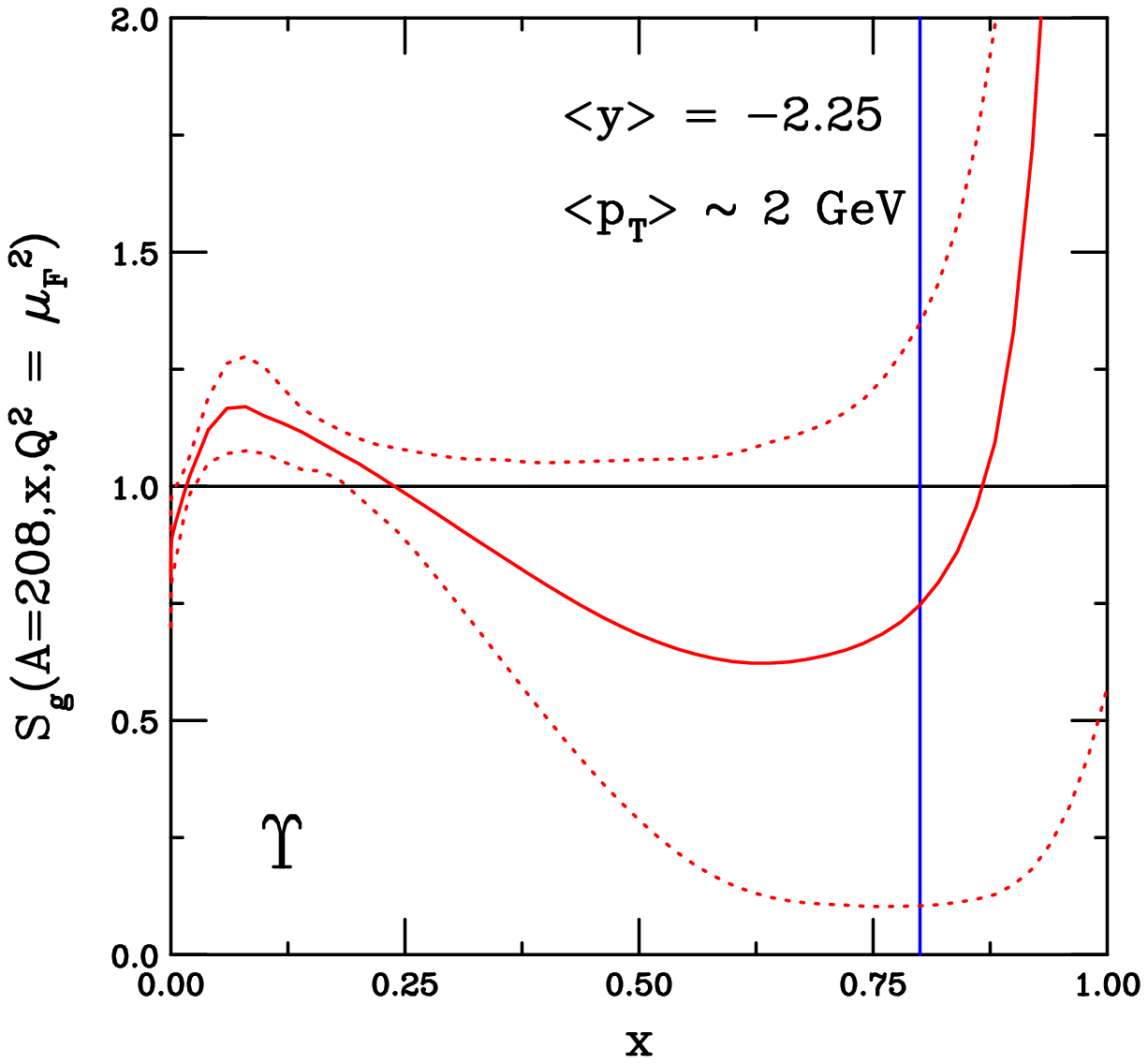} \\
\includegraphics[width=0.495\textwidth]{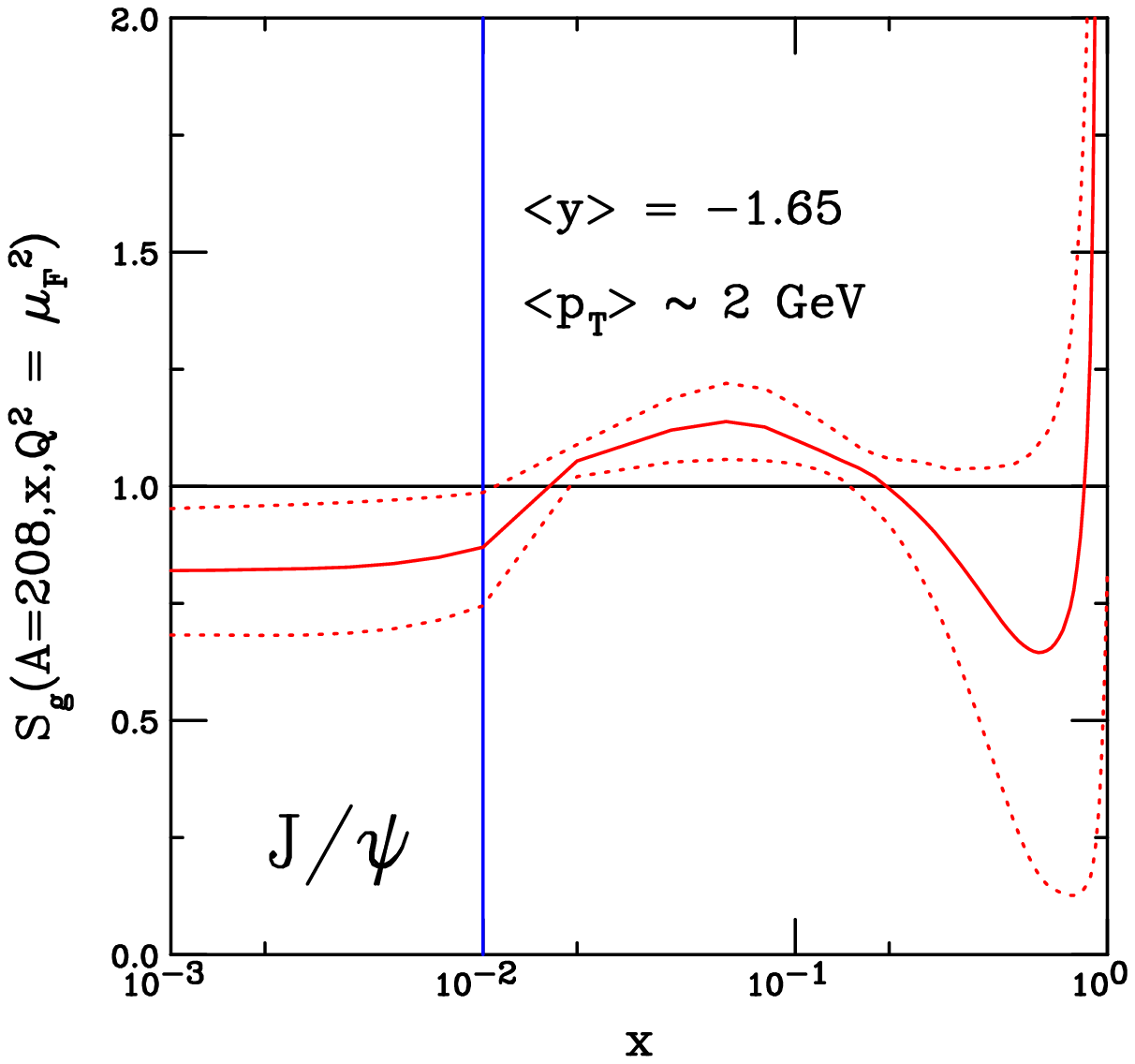}
\includegraphics[width=0.495\textwidth]{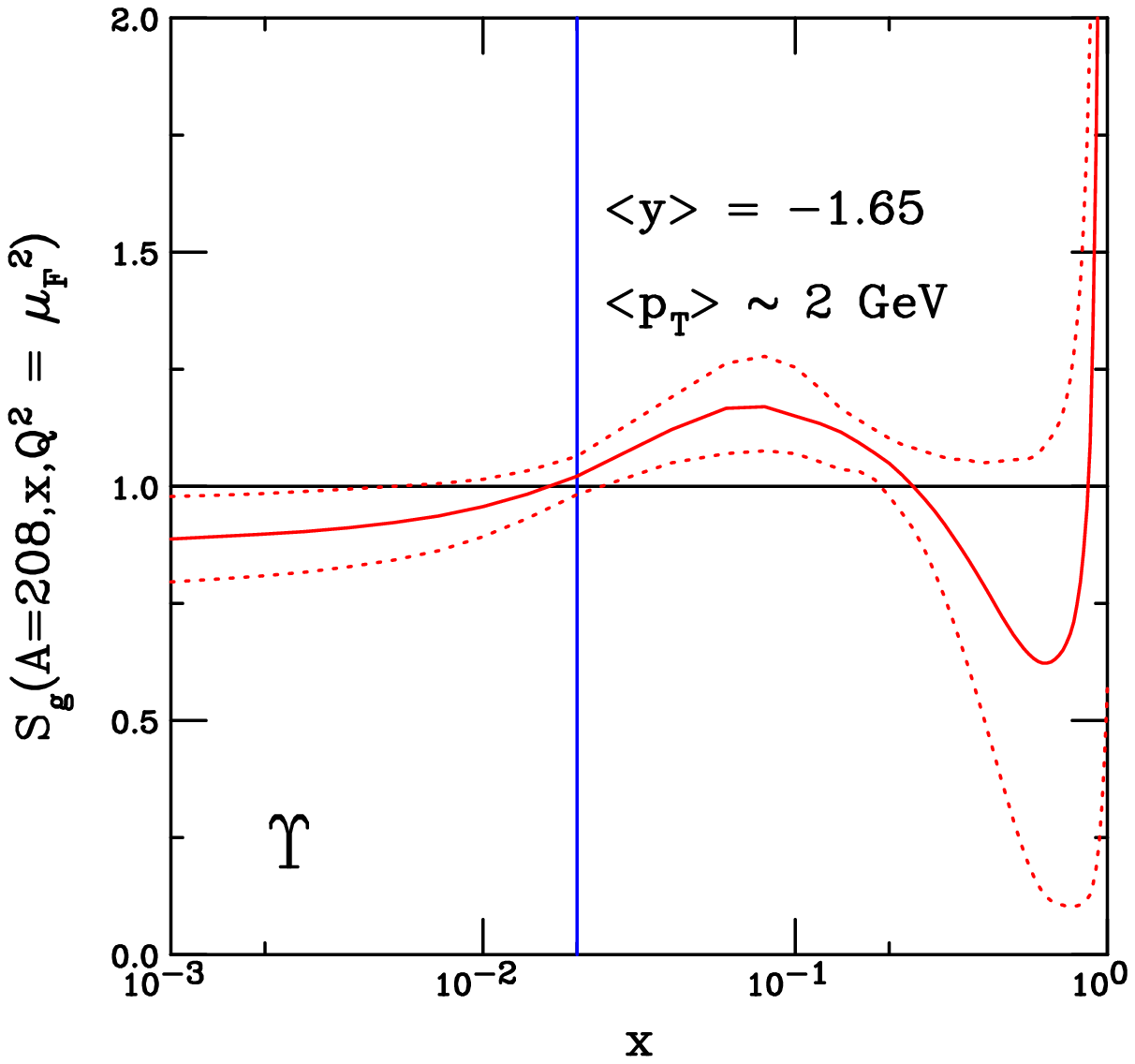} 
\end{center}
\caption{The EPS09 NLO 
shadowing ratios for $J/\psi$ (left) and $\Upsilon$(1S) (right)
production.  The solid curve in each plot is the central EPS09 NLO result
while the dotted curves outline the shadowing uncertainty band.
The upper plots are on a linear scale to emphasize the large $x$
region while the lower ones are on a logarithmic scale to expand the low $x$
region.  The approximate kinematic area of interest is indicated by the 
vertical line in each case.
}
\label{EPS09fig}
\end{figure*}

One possibility for the AFTER@LHC experiment is to use the LHCb detector, either
as is, with $2 < y_{\rm lab} < 5$, or an improved LHCb (LHCb+), with 
$1 < y_{\rm lab} < 5$.

In the fixed-target kinematics of AFTER@LHC, with a 7 TeV proton beam, the 
rapidity 
range is $\Delta y = 4.8$, corresponding to a center of mass rapidity coverage
of $-2.8 < y_{\rm cms} < 0.2$ for LHCb or $-3.8 < y_{\rm cms} < 0.2$ for LHCb+
in $p+$Pb collisions at $\sqrt{s_{_{NN}}} = 115$ GeV.  
In this case, the Pb nucleus is the target.  If
$x_1$ refers to the momentum fraction probed in the proton beam and $x_2$ is
the momentum fraction probed in the lead target, in these kinematics, the
negative rapidity means that $x_2$ is large, $x_2 > 0.1$.  This $x_2$ range
has not been explored since early nuclear deep-inelastic scattering (nDIS)
measurements
such as by the European Muon Collaboration \cite{EMC} and at SLAC \cite{SLAC}
and has never been explored by gluon-dominated processes such as quarkonium
production.  AFTER@LHC would be the first experiment to probe these kinematics
since most fixed-target configurations studying quarkonium have placed the 
detectors downstream where $x_1 > x_2$, as at the CERN SPS \cite{NA38,NA50} and
the Fermilab Tevatron \cite{E866}.  The only quarkonium experiment to 
measure part of this backward, large $x_2$ region was HERA-B with its foils
placed around the edges of the proton beam at HERA \cite{HERA-B}.

On the other hand, with a 2.76 TeV lead beam, the rapidity 
range is $\Delta y = 4.3$, corresponding to a center of mass rapidity coverage
of $-2.3 < y_{\rm cms} < 0.7$ for LHCb or $-3.3 < y_{\rm cms} < 0.7$ for LHCb+
in Pb$+p$ collisions at $\sqrt{s_{_{NN}}} = 72$ GeV.  
In this case, the proton is the
target, with $x_2$, and the lead beam is assigned $x_1$.  Thus, the nuclear
momentum fractions probed are moderate, $x_1 \sim 0.01$.  This $x$ region
has been well studied in nDIS experiments but,
again, not for final states dominated by initial-state gluons.

Global fits to the nuclear parton densities (nPDFs), such as EPS09, typically
include nuclear deep-inelastic scattering data ($F_2$ in $l+A$ and 
$dF_2/d\ln Q^2$), Drell-Yan data, and, more recently, RHIC data such as $\pi^0$
production \cite{Eskola:2009}.  The range over which DGLAP evolution can be
applied ($Q^2 > 1$ GeV$^2$) for fixed-target nDIS limits the minimum $x$
values probed.  In addition, such analyses do not take into account the
possibility of any other cold nuclear matter effects so that the possibility
of an effect such as initial-state energy loss in matter by quarks in Drell-Yan
dilepton production is folded in with the global analysis of nuclear shadowing.
Quarkonium production is particularly subject to other cold nuclear matter
effects such as energy loss in matter, break up of the quarkonium state by
nucleons (nuclear absorption), and interactions with comoving hadrons, see
e.g. Ref.~\cite{RV_e866} for a discussion.  For the purposes of this paper,
we focus only on the expected effects of shadowing.

Figure \ref{EPS09fig} shows the EPS09 NLO gluon shadowing parameterization as a 
function of momentum fraction, $x$.  The scales at which the results are
shown correspond to those used in the CEM for $J/\psi$ (left-hand side)
and $\Upsilon$ (right-hand side) production.  Along with the central set,
denoted by the solid curves, the dotted curves display the uncertainty band.
EPS09 obtains 30 additional sets of shadowing ratios by varying each of
the 15 parameters within one standard deviation of the mean.  The differences
are added in quadrature to produce the uncertainty band in the shadowing ratio
$S_g$.  (We note
that the uncertainties in the corresponding LO set are larger while the 
central shadowing effect is greater at LO than at NLO.  For more details
concerning this set as well as differences between other available nPDF sets,
see Ref.~\cite{RV_inprog}.)

The vertical blue line in each panel shows the average $x$ value for the
final quarkonium states at each energy, $p+$Pb on top and Pb$+p$ on the
bottom.  This is obtained by estimating the average $x$ value from the simpler
$2 \rightarrow 1$ kinematics of the LO CEM with 
$x_{1,2} = (2m/\sqrt{s_{_{NN}}})\exp(\pm y)$ and replacing $m$ by 
$m_T = \sqrt{m^2 + p_T^2}$ with $p_T^2 = 0.5(p_{T_Q}^2 + p_{T_{\overline Q}}^2)$.  
The $x$ value from the LO CEM represents a lower limit on $x$ relative to the
actual $2 \rightarrow 2$ and $2 \rightarrow 3$ kinematics of the LO and NLO 
contributions to the full NLO CEM calculation.  The average center of mass
rapidity, $\langle y_{\rm cms} \rangle$, shown on each panel is the 
approximate midpoint of our
chosen rapidity interval in each case.  The average $p_T$ of $\sim 2$ GeV
is near the peak of the $p_T$ distributions.  These values should not be
thought of as having the most weight in the actual calculations which
integrate over the rapidity interval for the $p_T$ distributions 
and all $p_T$ for the rapidity distributions.  Indeed, since the rapidity
distributions are steeply falling, the preponderance of the rate comes from the
upper end of the range in each case.  Thus the vertical lines represent
an estimate of the lower bound on the $x$ range at the given value of $p_T$.

The $p+$Pb kinematics emphasize high $x$ in the nucleus (top panels) and
thus explores an $x$ range rarely probed, especially by gluon-dominated 
processes.  It is partly in the 'EMC' region of the $x$ range and also moves
into the regime of 'Fermi motion', see a discussion of how the various
$x$ regions are parameterized by Eskola and collaborators in 
Refs.~\cite{EKS981,EKS982}.  Given the shortage of direct gluon-induced data
in the global analyses, the gluon shadowing ratios are constrained by the 
momentum sum rule.  The gluon shadowing ratios shown in the top half of 
Fig.~\ref{EPS09fig} are plotted on a linear scale to highlight the large $x$
region.  Here the scale dependence is very weak, illustrated by the
similarities in the results for the two quarkonium scales shown, while the 
uncertainties in the nPDF extraction are the largest.  AFTER@LHC measurements
could help narrow this uncertainty range.

On the other hand, the Pb$+p$ kinematics are in an $x$ region where 
quark-dominated processes, as in nDIS, are well measured and the uncertainties
can be expected to be relatively small.

\section{Cold nuclear matter effects and quarkonium production rates}
\label{rates}

There are other possible cold matter effects on $J/\psi$ production in
addition to that of shadowing: breakup
of the quarkonium state due to inelastic interactions with nucleons 
(absorption) or produced hadrons (comovers) and energy loss in cold matter. 
  
The quarkonium absorption cross section at midrapidity
was seen to decrease with center-of-mass energy
in Ref.~\cite{Lourenco:2008sk}, independent of whether shadowing effects 
were included or not.  It was also seen that incorporating shadowing into the
extraction of the absorption cross section required a larger effective
cross section \cite{Lourenco:2008sk}.  Extrapolating from
the results of Ref.~\cite{Lourenco:2008sk} to the energy range of AFTER@LHC, we
can expect an effective absorption cross section of a few millibarns at
midrapidity.  Away from midrapidity, the effective absorption cross section
was seen to rise at forward Feynman $x$, $x_F$ \cite{Lourenco:2008sk}, which
could be attributed to energy loss in matter \cite{ArleoPeigne}.  When a 
similar analysis was extended to the RHIC collider geometry, the effective
absorption cross section was also seen to increase in the backward region
\cite{McGlinchey:2012bp}.  Such behavior could be attributed to the quarkonium
state being fully formed inside the nucleus.  The $p+$Pb kinematics of AFTER@LHC
would be an ideal environment to study absorption in the target if other
observables can also probe shadowing effects to disentangle the two.

The shadowing results shown here are
obtained in the color evaporation model (CEM) \cite{Barger:1979js,Barger:1980mg}
at next-to-leading order in the
total cross section \cite{Gavai:1994in}.
In the CEM, the quarkonium 
production cross section is some fraction, $F_C$, of 
all $Q \overline Q$ pairs below the $H \overline H$ threshold where $H$ is
the lowest mass heavy-flavor hadron,
\begin{eqnarray}
\sigma_C^{\rm CEM}(s) & = & F_C \sum_{i,j} 
\int_{4m^2}^{4m_H^2} ds
\int dx_1 \, dx_2~ \nonumber \\  &  & f_i^p(x_1,\mu_F^2)~ f_j^p(x_2,\mu_F^2)~ 
\hat\sigma_{ij}(\hat{s},\mu_F^2, \mu_R^2) \, 
\, , \label{sigtil}
\end{eqnarray} 
where $ij = q \overline q$ or $gg$ and $\hat\sigma_{ij}(\hat s)$ is the
$ij\rightarrow Q\overline Q$ subprocess cross section.    
The normalization factor $F_C$ is fit 
to the forward (integrated over $x_F > 0$) 
$J/\psi$ cross section data and the combined $\Upsilon$($n$S) state data at
midrapidity.  We use the code of Ref.~\cite{MNRcode} with the mass cut
implemented.

The same values of the central charm quark
mass and scale parameters are employed as those 
found for open charm, $m_c = 1.27 \pm 0.09$~GeV,
$\mu_F/m_c = 2.10 ^{+2.55}_{-0.85}$, and $\mu_R/m_c = 1.60 ^{+0.11}_{-0.12}$ 
\cite{Nelson:2012bc}.
The normalization $F_C$ is obtained for the central set,
$(m_c,\mu_F/m_c, \mu_R/m_c) = (1.27 \, {\rm GeV}, 2.1,1.6)$.  
The calculations for the extent of the mass and scale uncertainties are
multiplied by the same value of $F_C$ to
obtain the extent of the $J/\psi$ uncertainty band \cite{Nelson:2012bc}.
These values reproduce the energy dependence of $J/\psi$ production from
fixed-target to collider energies.  The resulting rapidity and $p_T$
distributions also agree with the $p+p$ data from RHIC and the LHC at $\sqrt{s}
=200$ GeV and 7 TeV respectively \cite{Nelson:2012bc}.

We calculate $\Upsilon$ production in the same manner, with the central
result obtained for $(m_b,\mu_F/m_b, \mu_R/m_b) = (4.65 \, {\rm GeV}, 1.4,1.1)$
\cite{Nelson_inprog}.  We have also found good agreement with the $\sqrt{s}$,
and $p_T$ distributions from previous measurements \cite{Nelson_inprog}.
Unfortunately, the uncertainties from RHIC measurements are rather large and 
few data are available on the shape of the $\Upsilon$ rapidity distributions.

To obtain the quarkonium $p_T$ distributions at low $p_T$, intrinsic transverse 
momentum, $k_T$, smearing for quarkonium
is included in the initial-state parton densities \cite{SchulerV}.
Since the MNR code cancels divergences numerically, instead of slowing the
calculations by adding more integrations, the $k_T$ kick is added in the
final, rather than the initial, state \cite{MNRcode}. 
The Gaussian function 
$g_p(k_T) = \pi \langle k_T^2 \rangle_p^{-1} 
\exp(-k_T^2/\langle k_T^2 \rangle_p)$ \cite{MLM1}, 
multiplies the parton distribution functions for both hadrons, 
assuming the $x$ and $k_T$ dependencies in the initial partons completely
factorize.  If factorization applies, 
it does not matter whether the $k_T$ dependence
appears in the initial or final state if the kick is not too large. 
The effect of the intrinsic $k_T$ on the shape of the $J/\psi$ $p_T$ 
distribution can be expected to decrease as 
$\sqrt{s}$ increases because the average $p_T$ of the $J/\psi$ also increases 
with energy.  However, the value of $\langle k_T^2 \rangle$ may increase with
$\sqrt{s}$.  
We can check the energy dependence of $\langle k_T^2 \rangle$ by
the shape of the $J/\psi$ $p_T$ distributions at central and forward rapidity
at RHIC.  We find that 
$\langle k_T^2 \rangle = 1 + (1/12)\ln(\sqrt{s}/20) \approx 1.19$~GeV$^2$ 
at $\sqrt{s} = 200$ GeV agrees well with the $J/\psi$ $p_T$ 
distributions \cite{Nelson:2012bc}.
All the 
calculations are NLO in the total cross section and assume that the intrinsic
$k_T$ broadening is the same in $p+p$ as in $p+$Pb.  While the broadening is
expected to increase in collisions with nuclei as projectile, target or both,
the agreement of the $J/\psi$ $p+$Pb ratio $R_{p{\rm Pb}}(p_T)$ with the LHC 
data is better without any additional broadening \cite{RV_inprog}.
Therefore we do not change the value here.

\subsection{$J/\psi$ and $\Upsilon$ (1S) production in $p+$Pb collisions at
$\sqrt{s_{_{NN}}} = 115$ GeV}

\begin{figure*}[thp!]
\begin{center}
\includegraphics[width=0.495\textwidth]{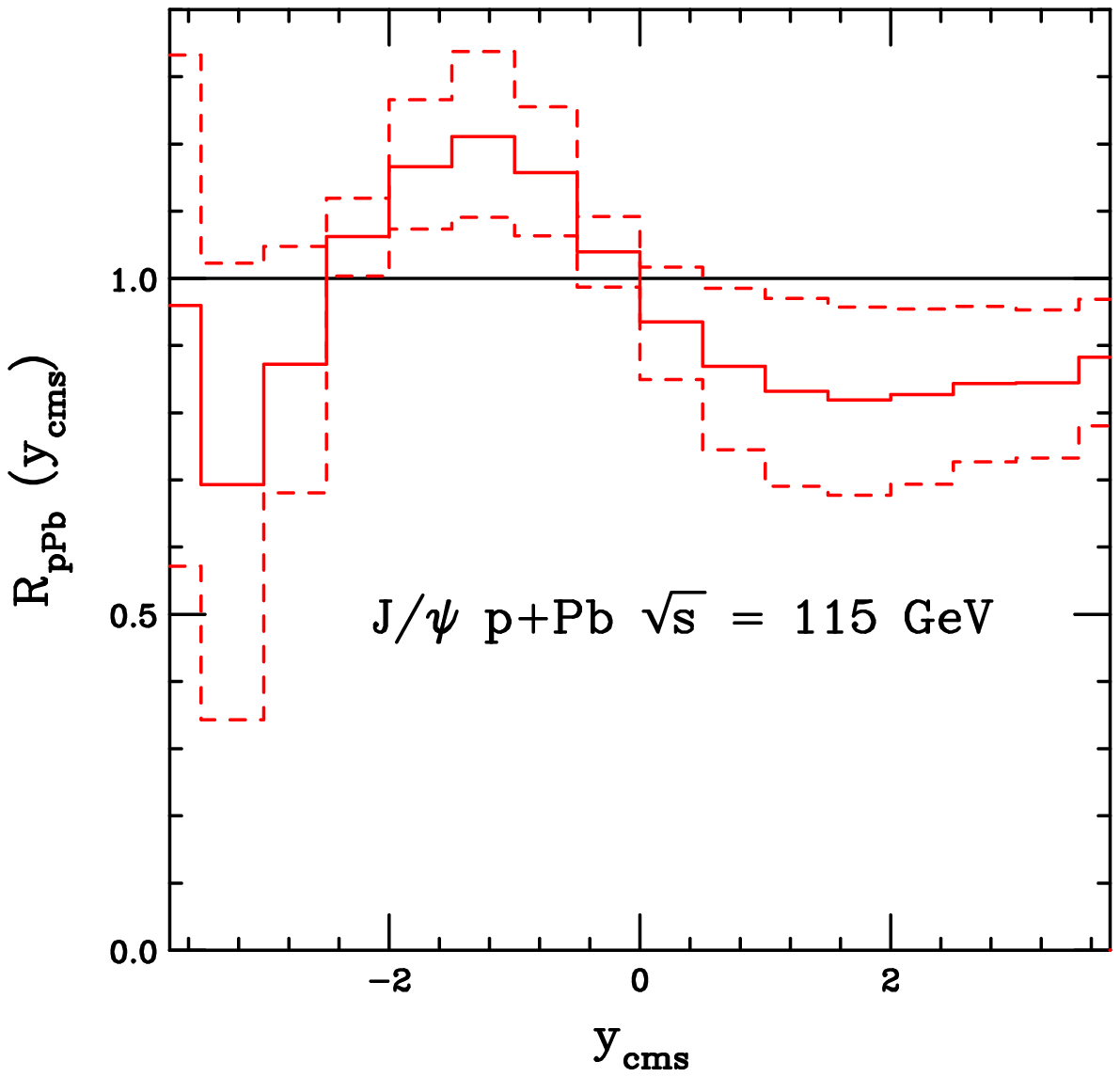}
\includegraphics[width=0.495\textwidth]{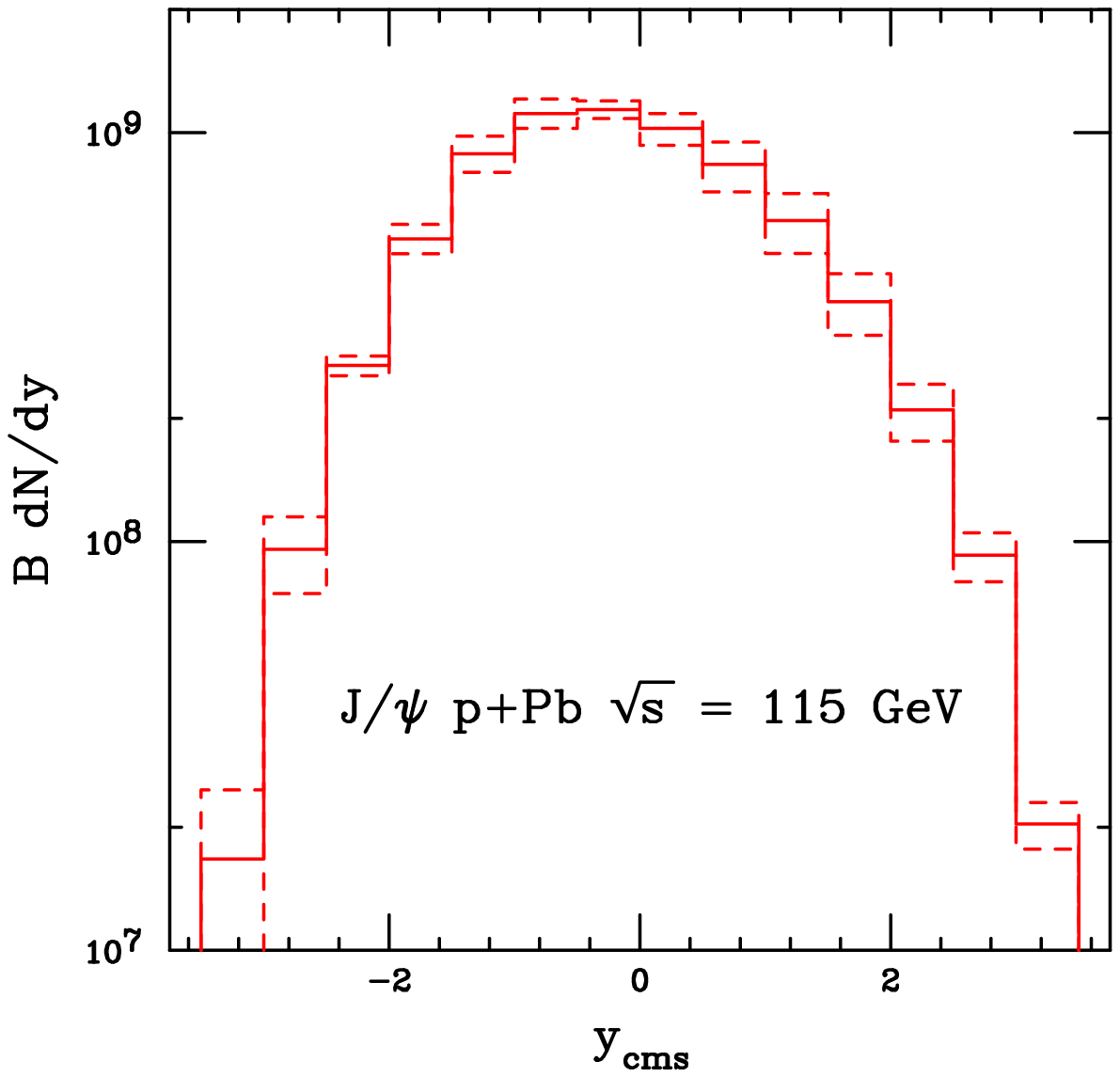} \\
\includegraphics[width=0.495\textwidth]{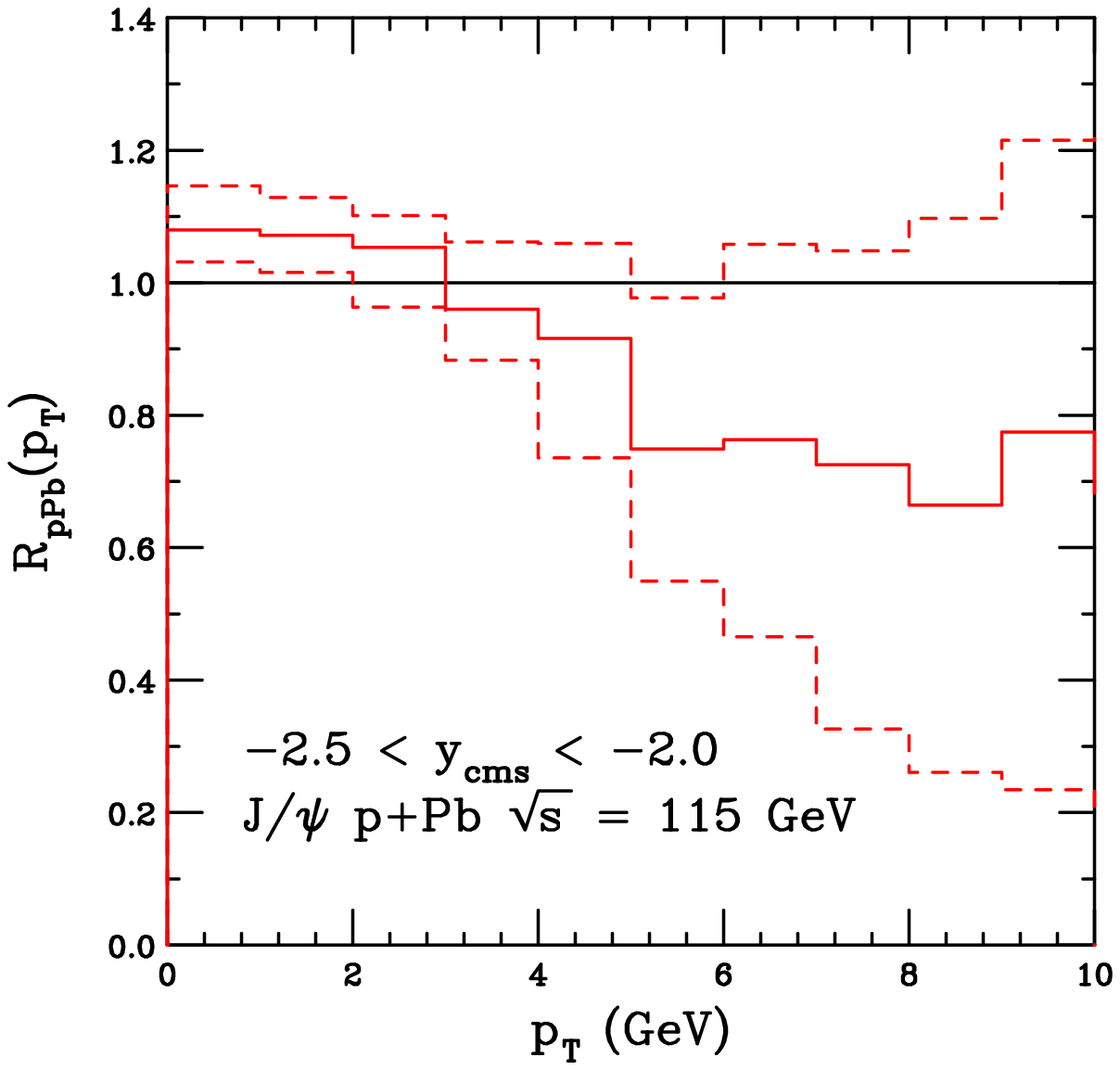}
\includegraphics[width=0.495\textwidth]{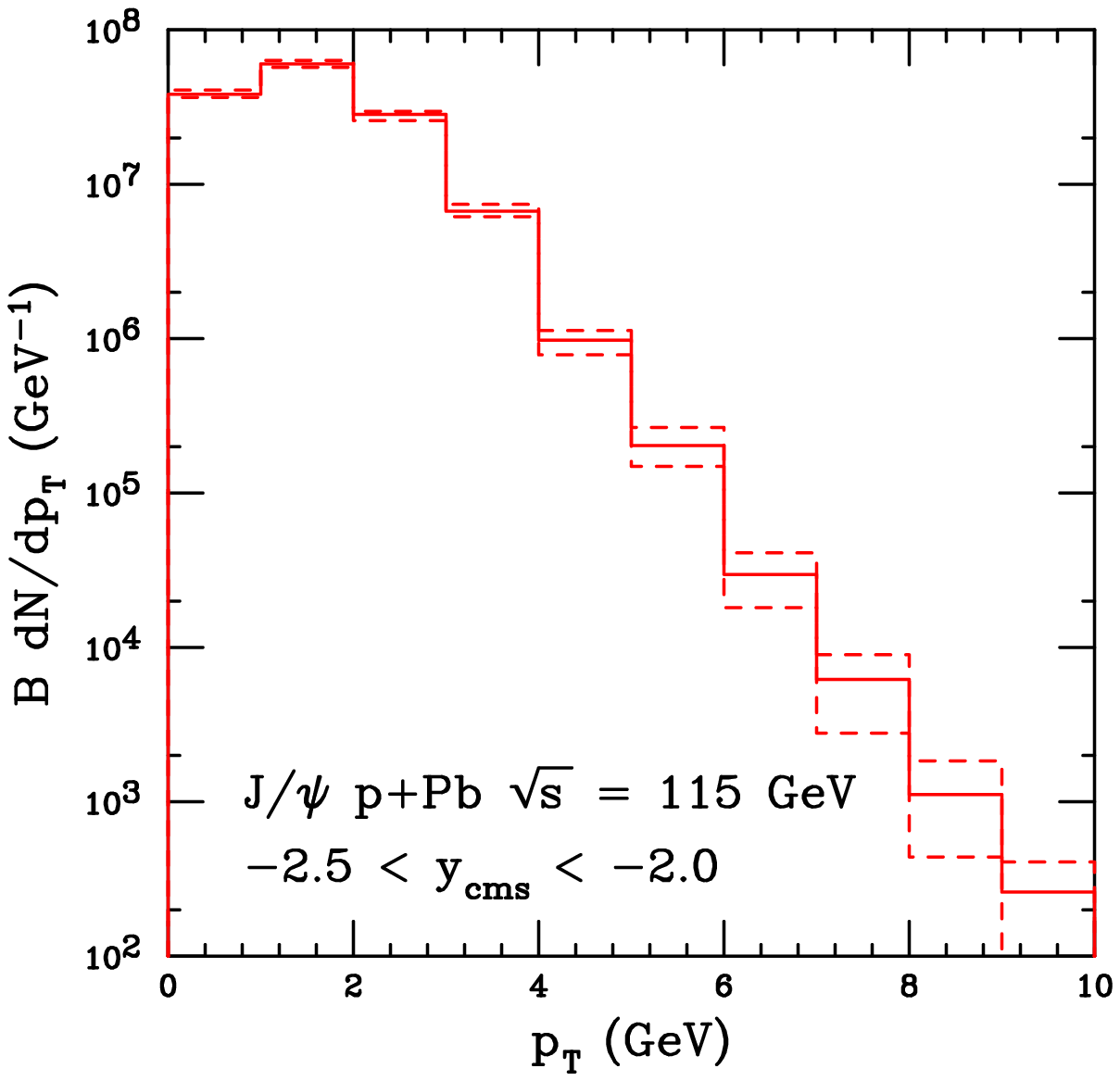} 
\end{center}
\caption{The predicted $J/\psi$ shadowing ratios (left) and rates (right)
as a function of center of mass rapidity (top) and $p_T$ (bottom) 
for $p+$Pb collisions at $\sqrt{s_{_{NN}}} = 115$ GeV.  The solid curve in 
each plot is the central EPS09 NLO result
while the dotted curves outline the shadowing uncertainty band.
}
\label{Jpsi_pPb}
\end{figure*}

\begin{figure*}[thp!]
\begin{center}
\includegraphics[width=0.495\textwidth]{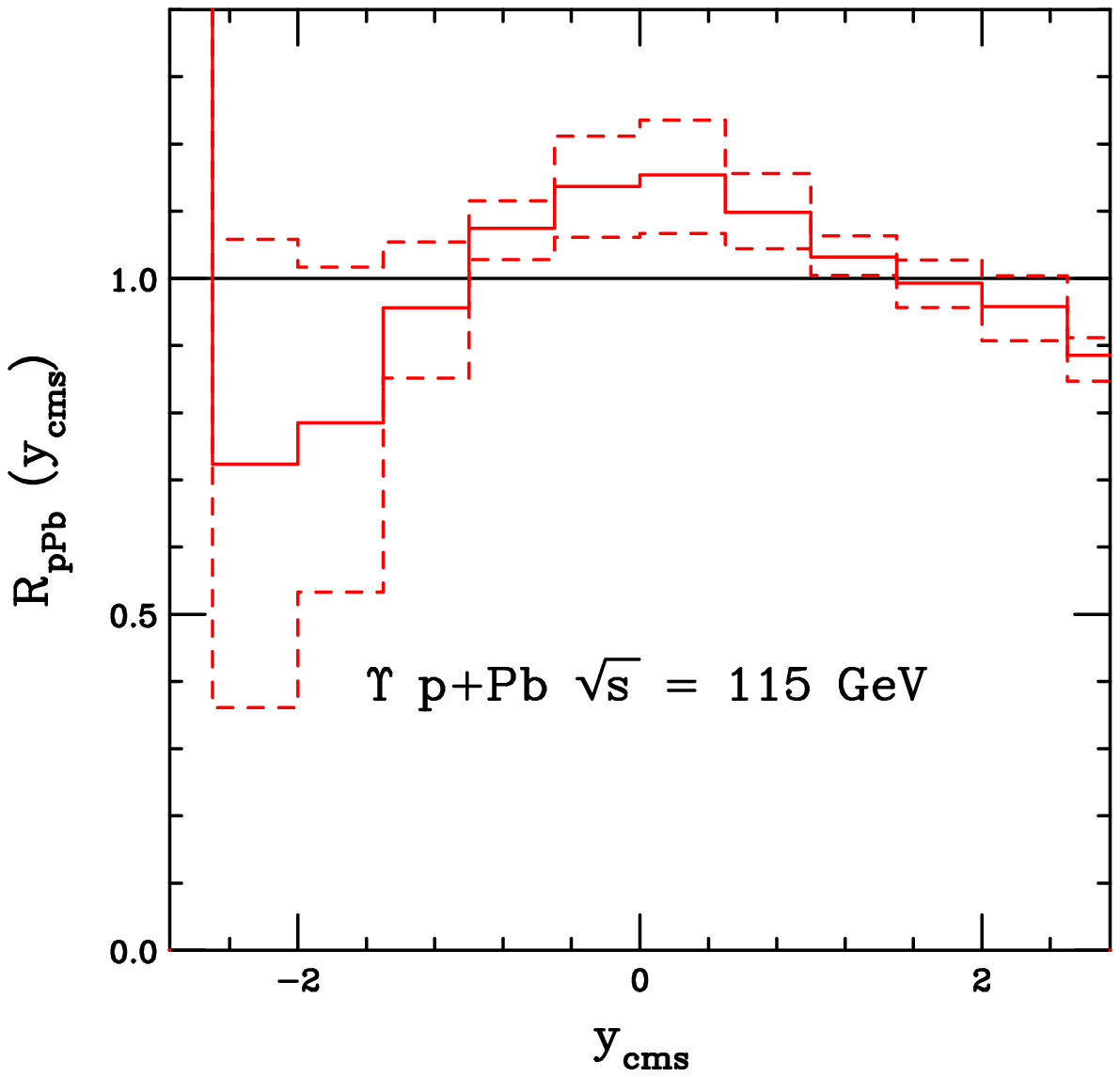}
\includegraphics[width=0.495\textwidth]{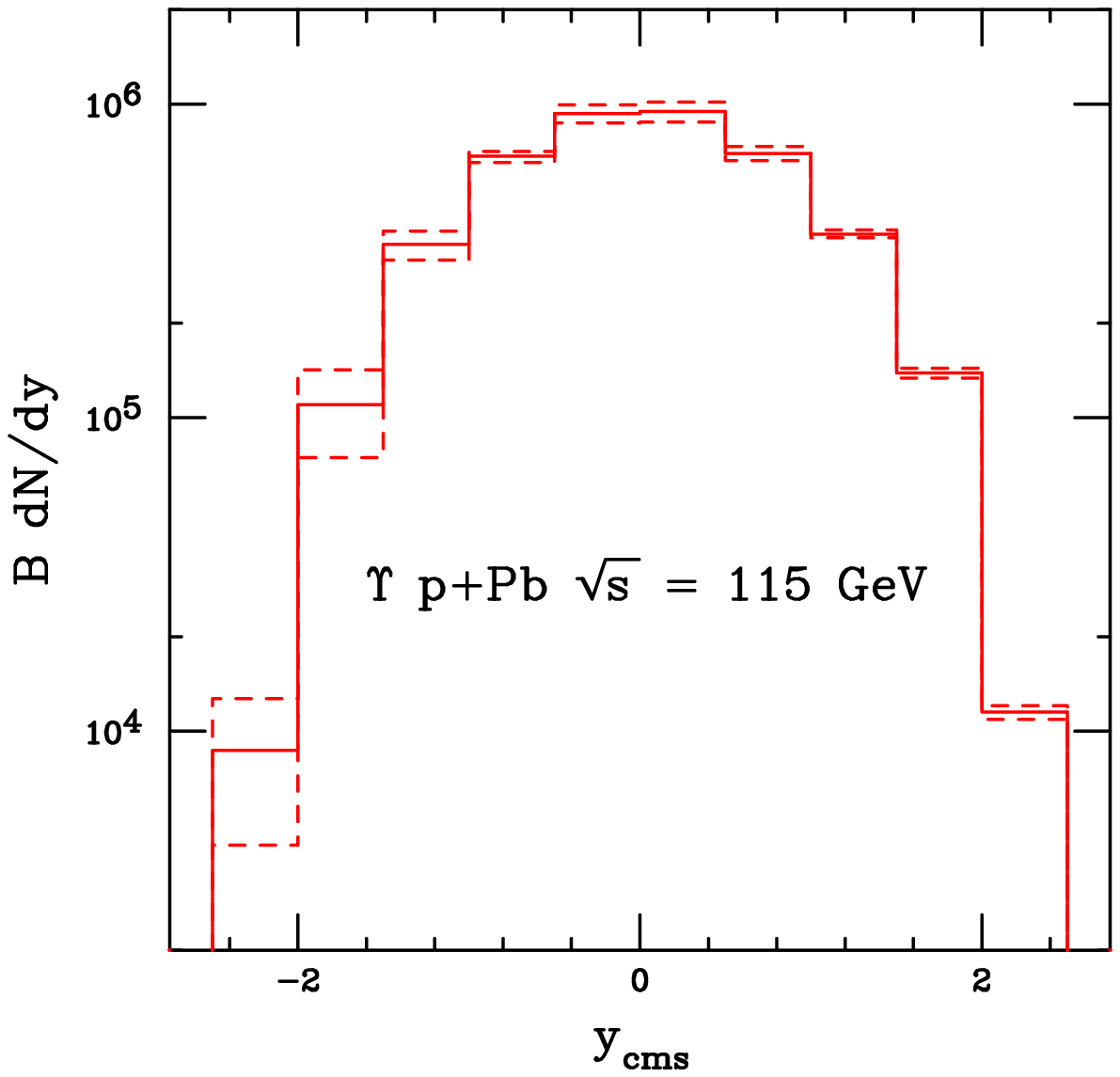} \\
\includegraphics[width=0.495\textwidth]{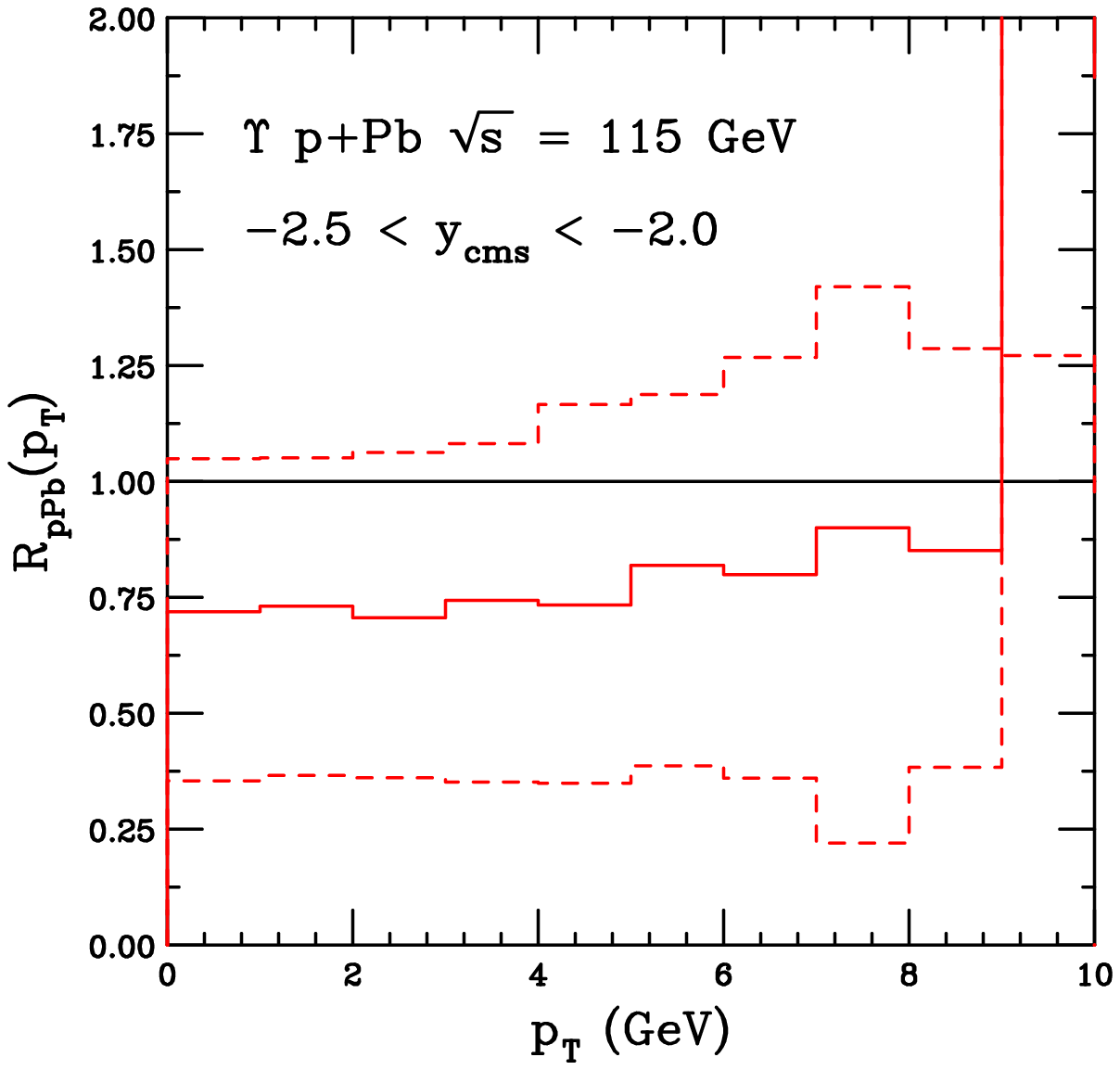}
\includegraphics[width=0.495\textwidth]{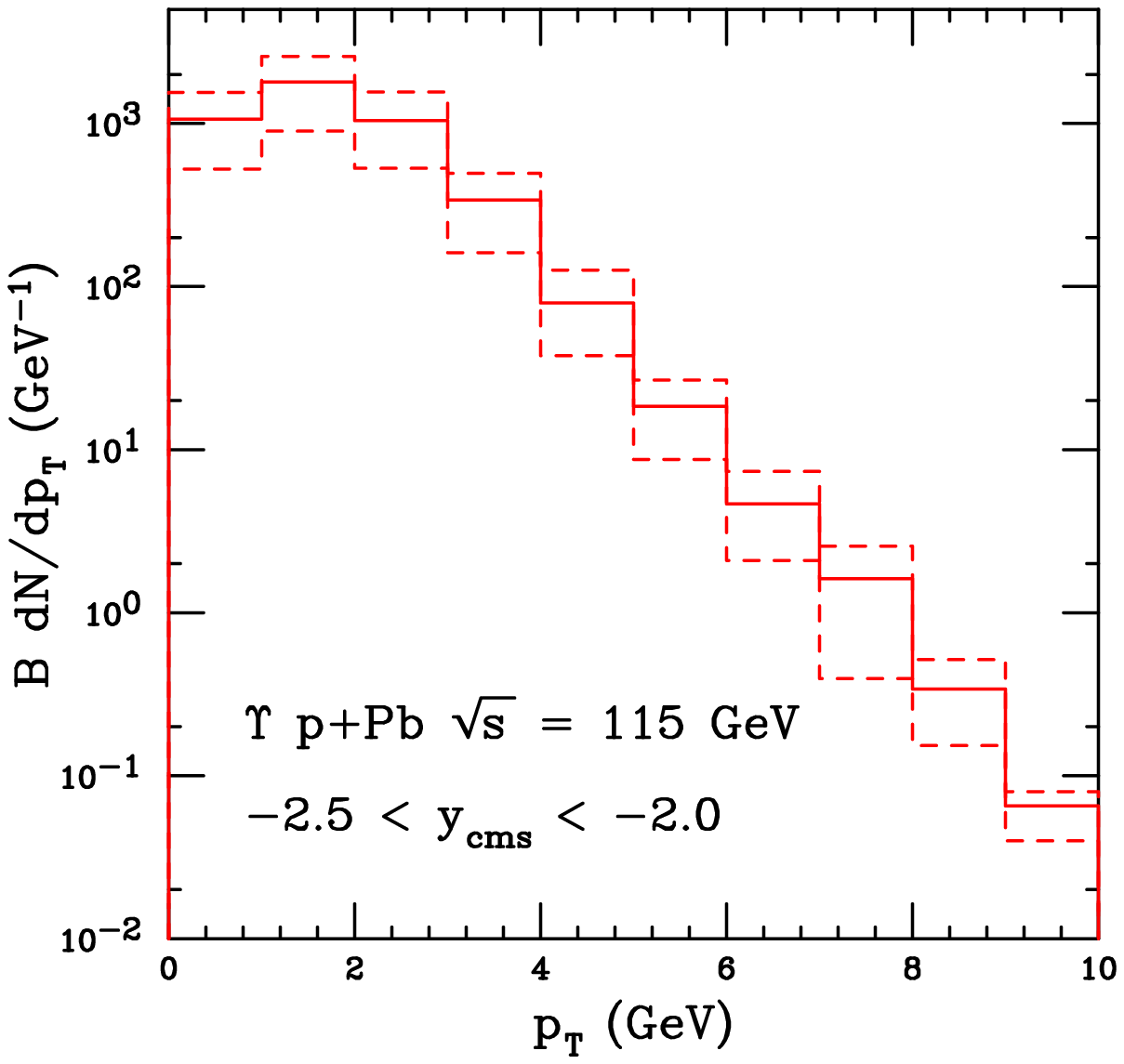} 
\end{center}
\caption{The predicted $\Upsilon$(1S) shadowing ratios (left) and rates (right)
as a function of center of mass rapidity (top) 
and $p_T$ (bottom) for $p+$Pb collisions
at $\sqrt{s_{_{NN}}} = 115$ GeV.  The solid curve in each plot is the 
central EPS09 NLO result
while the dotted curves outline the shadowing uncertainty band.
}
\label{Ups_pPb}
\end{figure*}

In this section, the results for $J/\psi$ and $\Upsilon$ shadowing in $p+$Pb
collisions at $\sqrt{s_{_{NN}}} = 115$ GeV are presented.  Figure~\ref{Jpsi_pPb}
shows the results for $J/\psi$ while Fig.~\ref{Ups_pPb} shows the $\Upsilon$
results.  In both cases, the left-hand side shows the ratios 
$R_{p{\rm Pb}}$ as a function of $y$ (top) and $p_T$ in the rapidity range
$-2.5 < y_{\rm cms} < -2$ (bottom).  The rates to
dileptons in the rapidity acceptance,
assuming a lead target, are shown on the right-hand side of the 
figures.  

In the kinematics of this configuration, the large $x_2$ in the
nucleus puts the peak for Fermi motion at the most negative rapidities.  
(The full center-of-mass rapidity range for $J/\psi$'s produced at this energy 
is $|y_{\rm cms}| = \ln(\sqrt{s}/m) < 3.8$ for the mass and scale parameters
appropriate for the CEM calculation.)
The
EMC region is in the $J/\psi$ rapidity acceptance.  There is a steep drop in
$R_{p{\rm Pb}}(y_{\rm cms})$ as $y_{\rm cms}$ decreases 
from $-2$ to $-2.5$, changing the central
value of $R_{p{\rm Pb}}(y_{\rm cms})$ by $\sim 30$\% over the range.  
The decrease into the
EMC region is more apparent as a function of $p_T$ where the region is expanded
for $p_T < 10$ GeV.  The large uncertainty in this $x$ range, as emphasized
in the upper plots of Fig.~\ref{EPS09fig}, is enhanced here.

The rates as a function of rapidity for $J/\psi$ and $\Upsilon$(1S) decays to
lepton pairs are shown in Table~\ref{RateTable}.  While the rates are shown for
the entire rapidity range, the broad LHCb+ center of mass rapidity
acceptance ends at $y_{\rm cms} \sim 1$.  The rates are given in bins of
$\Delta y_{\rm cms} = 0.5$ with the value of $y_{\rm cms}$ at the center of the
bin shown in Table~\ref{RateTable}.  The rates 
include the branching ratios to lepton pairs.

The $J/\psi$ rates in $p+$Pb collisions are very high.
The rate is $B dN/dy \sim 2.7 \times 10^8$ in the chosen
backward rapidity bin of $-2.5 < y_{\rm cms} < -2$, see the upper right panel
of Fig.~\ref{Jpsi_pPb}.  The cross section is
rather high since $\sqrt{s_{_{NN}}}$ 
is above the region where the 
production cross section is still increasing
steeply with $\sqrt{s_{_{NN}}}$.  In addition, the $J/\psi$ production range in 
rapidity is fully within the AFTER@LHC acceptance.

Finally, even though the rates fall off quickly with $p_T$, more than 
100 events can be collected at $p_T \sim 9.5$ GeV, see the lower right panel
of Fig.~\ref{Jpsi_pPb} and the upper part of Table~\ref{RateTable_pT}, likely 
enough to determine where 
$R_{p{\rm Pb}}(p_T)$ lies within the EPS09 band.

\begin{table}[htbp]
\begin{center}
\begin{tabular}{|l|r|c|c|}\hline
System & $y_{\rm cms}$ & $N(J/\psi \rightarrow l^+ l^-)$ & 
$N(\Upsilon({\rm 1S}) \rightarrow l^+ l^-)$ \\ \hline
                     & -3.75 & $2.32 \times 10^5$  & -  \\
                     & -3.25 & $1.67 \times 10^7$  & -  \\
                     & -2.75 & $9.56 \times 10^7$  & -  \\
                     & -2.25 & $2.69 \times 10^8$  & $8.68 \times 10^3$ \\
                     & -1.75 & $5.50 \times 10^8$  & $1.10 \times 10^5$ \\
                     & -1.25 & $8.88 \times 10^8$  & $3.56 \times 10^5$ \\
                     & -0.75 & $1.11 \times 10^9$  & $6.81 \times 10^5$ \\
$p+$Pb               & -0.25 & $1.14 \times 10^9$  & $9.33 \times 10^5$ \\
$\sqrt{s_{_{NN}}} = 115$ GeV 
                     &  0.25 & $1.02 \times 10^9$  & $9.47 \times 10^5$ \\ 
                     &  0.75 & $8.36 \times 10^8$  & $6.96 \times 10^5$ \\
                     &  1.25 & $6.10 \times 10^8$  & $3.85 \times 10^5$ \\
                     &  1.75 & $3.86 \times 10^8$  & $1.39 \times 10^5$ \\
                     &  2.25 & $2.10 \times 10^8$  & $1.15 \times 10^4$ \\
                     &  2.75 & $9.25 \times 10^7$  & - \\
                     &  3.25 & $2.04 \times 10^7$  & - \\
                     &  3.75 & $2.13 \times 10^5$  & - \\ \hline \hline
                     & -3.25 & $1.54 \times 10^3$  & - \\
                     & -2.75 & $1.05 \times 10^5$  & - \\
                     & -2.25 & $4.36 \times 10^5$  & - \\
                     & -1.75 & $9.67 \times 10^5$  & $5.36 \times 10^1$ \\
                     & -1.25 & $1.78 \times 10^6$  & $5.10 \times 10^2$ \\
                     & -0.75 & $2.79 \times 10^6$  & $1.20 \times 10^3$ \\
Pb$+p$               & -0.25 & $3.72 \times 10^6$  & $1.66 \times 10^3$ \\
$\sqrt{s_{_{NN}}} = 72$ GeV  
                     &  0.25 & $4.15 \times 10^6$  & $1.59 \times 10^3$ \\
                     &  0.75 & $3.64 \times 10^6$  & $1.03 \times 10^3$ \\ 
                     &  1.25 & $2.42 \times 10^6$  & $3.78 \times 10^2$ \\
                     &  1.75 & $1.24 \times 10^6$  & $3.77 \times 10^1$ \\
                     &  2.25 & $4.58 \times 10^5$  & - \\
                     &  2.75 & $8.68 \times 10^4$  & - \\
                     &  3.25 & $1.55 \times 10^3$  & - \\  \hline
\end{tabular}
\end{center}
\caption[]{The rates per 0.5 unit rapidity for $J/\psi$ and $\Upsilon$(1S) in
the two scenarios discussed in the text.  The values are given for the 
EPS09 NLO central set. }
\label{RateTable}
\end{table}

\begin{table}[htbp]
\begin{center}
\begin{tabular}{|l|r|c|c|}\hline
System & $p_T$ (GeV) & $N(J/\psi \rightarrow l^+ l^-)$ & 
$N(\Upsilon ({\rm 1S}) \rightarrow l^+ l^-)$ \\ \hline
                     & 0.5  & $3.83 \times 10^7$    & $1.06 \times 10^3$ \\
                     & 1.5  & $6.62 \times 10^7$    & $1.79 \times 10^3$ \\
                     & 2.5  & $2.83 \times 10^7$    & $1.04 \times 10^3$ \\
                     & 3.5  & $6.69 \times 10^6$    & $3.40 \times 10^2$ \\
                     & 4.5  & $9.78 \times 10^5$    & $7.93 \times 10^1$ \\
$p+$Pb               & 5.5  & $2.03 \times 10^5$    & $1.84 \times 10^1$ \\
$\sqrt{s_{_{NN}}} = 115$ GeV 
                     & 6.5  & $2.96 \times 10^4$    & $4.64 \times 10^0$ \\
                     & 7.5  & $6.20 \times 10^3$    & $1.62 \times 10^0$ \\
                     & 8.5  & $1.12 \times 10^3$    & $3.40 \times 10^{-1}$ \\ 
                     & 9.5  & $2.60 \times 10^2$    & $6.51 \times 10^{-2}$ \\
                     & 10.5 & $3.96 \times 10^1$    & $1.03 \times 10^{-2}$ \\
\hline \hline
                     & 0.5  & $1.38 \times 10^5$    & $1.39 \times 10^1$ \\
                     & 1.5  & $2.36 \times 10^5$    & $2.26 \times 10^1$ \\
                     & 2.5  & $1.33 \times 10^5$    & $1.20 \times 10^1$ \\
                     & 3.5  & $3.90 \times 10^4$    & $3.39 \times 10^0$ \\
                     & 4.5  & $7.69 \times 10^3$    & $7.44 \times 10^{-1}$ \\
Pb$+p$               & 5.5  & $1.16 \times 10^3$    & $1.71 \times 10^{-1}$ \\
$\sqrt{s_{_{NN}}} = 72$ GeV  
                     & 6.5  & $2.11 \times 10^2$    & $4.69 \times 10^{-2}$ \\
                     & 7.5  & $4.15 \times 10^1$    & $1.28 \times 10^{-2}$ \\
                     & 8.5  & $8.62 \times 10^0$    & $3.25 \times 10^{-3}$ \\ 
                     & 9.5  & $1.89 \times 10^0$    & $6.72 \times 10^{-4}$ \\
                     & 10.5 & $3.39 \times 10^{-1}$  & $1.65 \times 10^{-4}$ \\
\hline
\end{tabular}
\end{center}
\caption[]{The $p_T$-dependent rates per 1 GeV $p_T$ bin for $J/\psi$ 
and $\Upsilon$(1S) in
the two scenarios discussed in the text.  The values are given for the 
EPS09 NLO central set. }
\label{RateTable_pT}
\end{table}

The $\Upsilon$(1S) rates, shown in Fig.~\ref{Ups_pPb}, are significantly 
lower.  At this
energy, the production cross section is still increasing rapidly so that the
the available phase space for
$\Upsilon$ production, $|y_{\rm cms}|<2.9$ in the CEM calculation.  Thus the 
AFTER@LHC acceptance is just inside the lower end of this range and
the rate for $\Upsilon$ production 
in this region is relatively low.
While the rates over all phase space can
be quite high, with nearly $10^6$ events at midrapidity, there are less
than $10^4$ events in the region $-2.5 < y_{\rm cms} < -2$, see 
Table~\ref{RateTable} and the lower right panel of Fig.~\ref{Ups_pPb}.

As shown in Table~\ref{RateTable_pT}, the $\Upsilon$ states that are produced
in the AFTER@LHC acceptance are primarily at low $p_T$, $p_T \leq 3$ GeV.
Indeed, there are fewer than 10 events per year for $p_T > 6$ GeV so that any
division into $p_T$ bins for $p_T > 5$ GeV is unlikely to be feasible.

The AFTER@LHC rapidity bin is in 
the EMC region and touching on the Fermi motion region at $y_{\rm cms} \sim -2.5$,
as seen on the upper left panel of Fig.~\ref{Ups_pPb}.  
The $p_T$ dependent ratio reflects the large uncertainty of the EMC region and
is almost independent of $p_T$ until $p_T \sim 9$ GeV in the EPS09
parameterization where it increases sharply. The low rate will make it 
difficult to study this interesting region in detail.

\subsection{$J/\psi$ and $\Upsilon$(1S) production in Pb+$p$ collisions at
$\sqrt{s_{_{NN}}} = 72$ GeV}

\begin{figure*}[thp!]
\begin{center}
\includegraphics[width=0.495\textwidth]{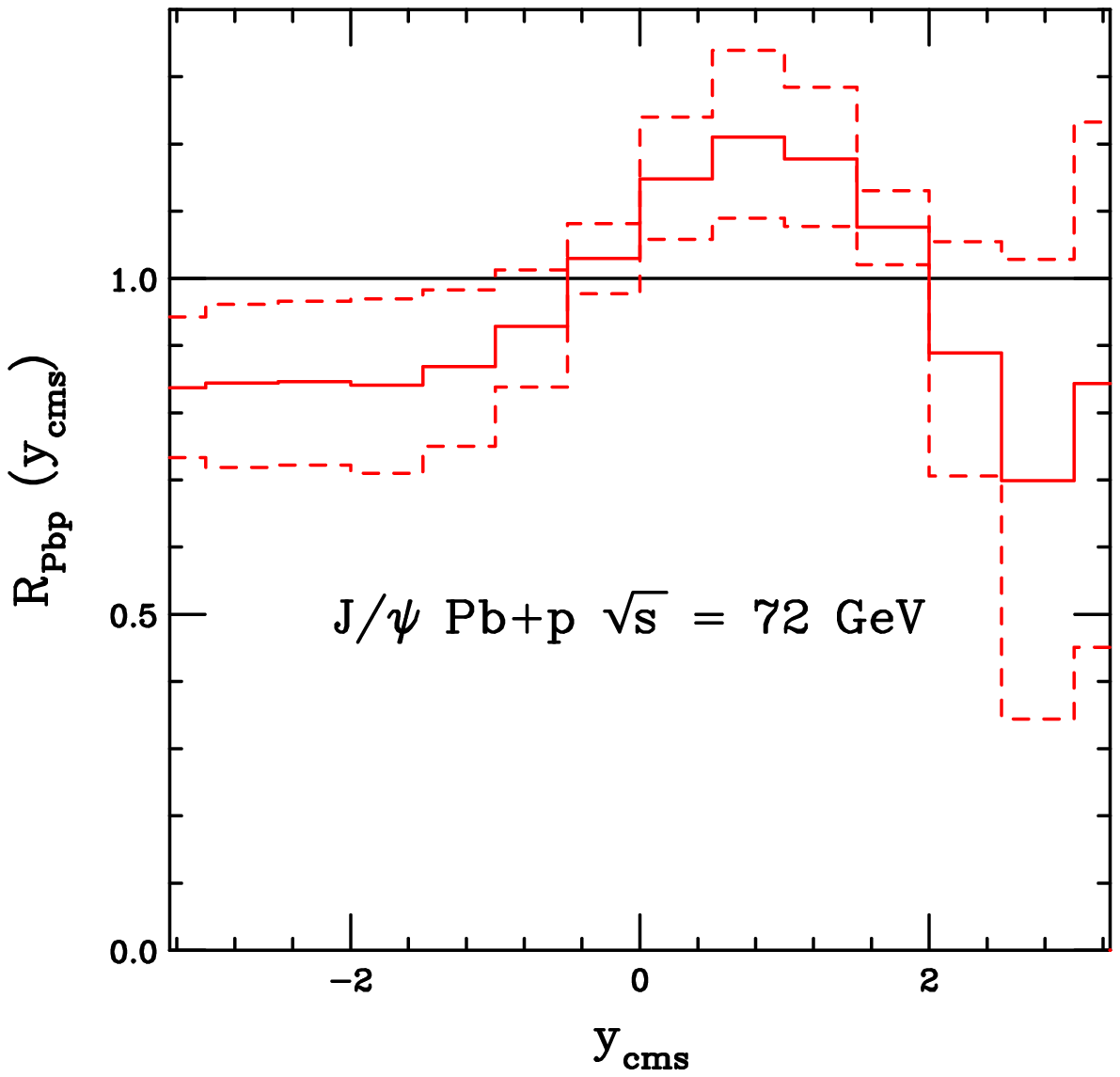}
\includegraphics[width=0.495\textwidth]{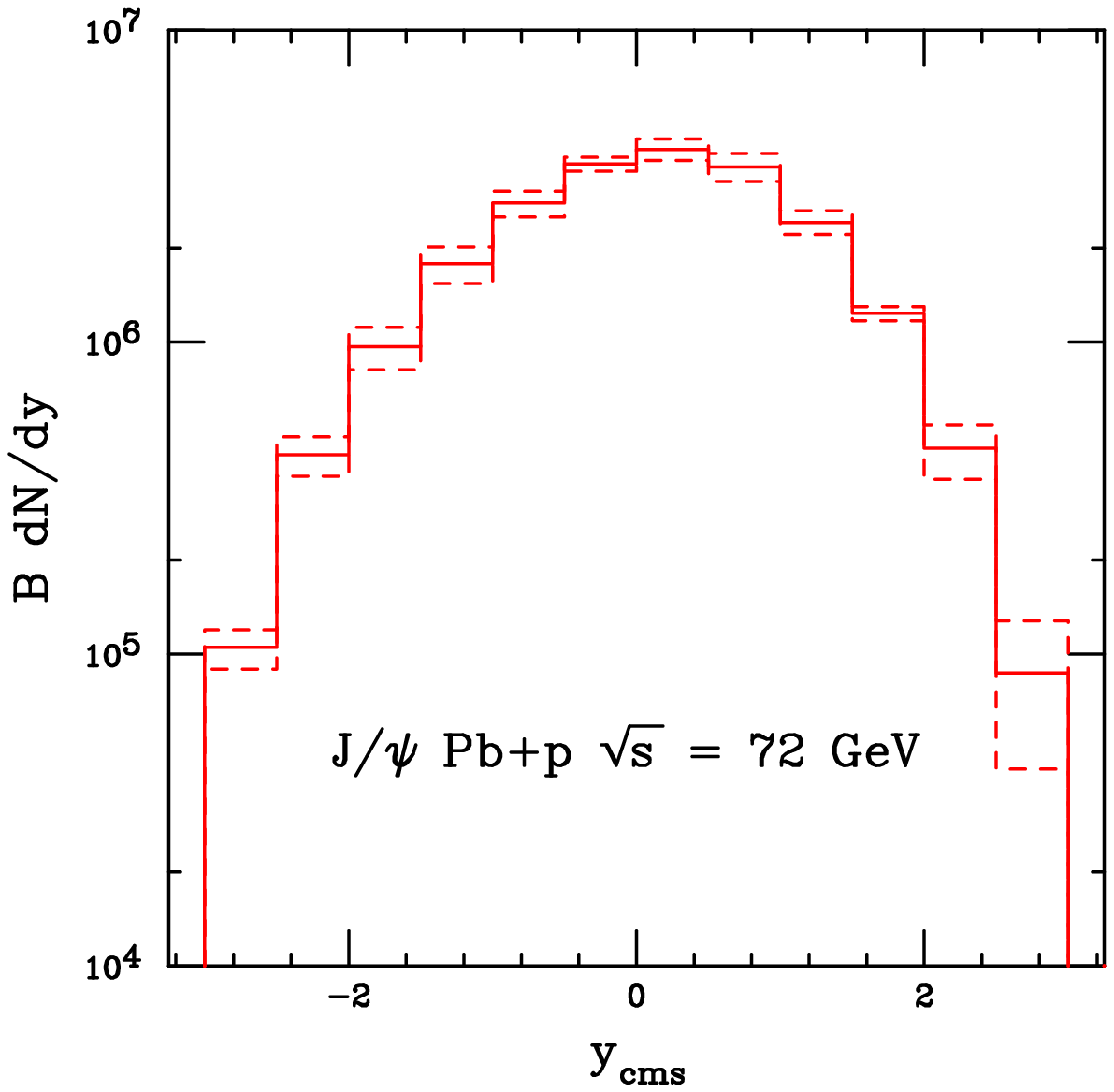} \\
\includegraphics[width=0.495\textwidth]{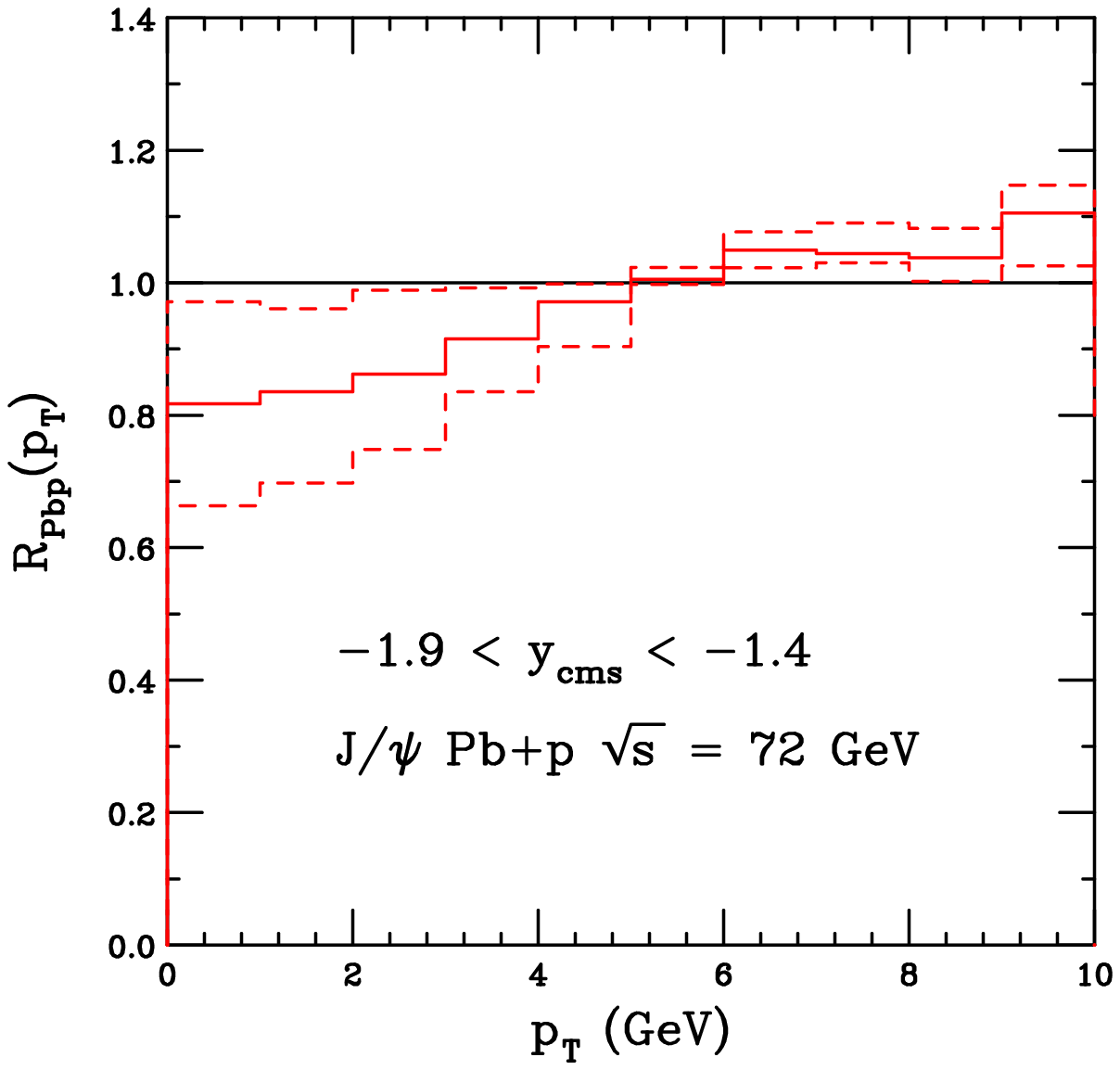}
\includegraphics[width=0.495\textwidth]{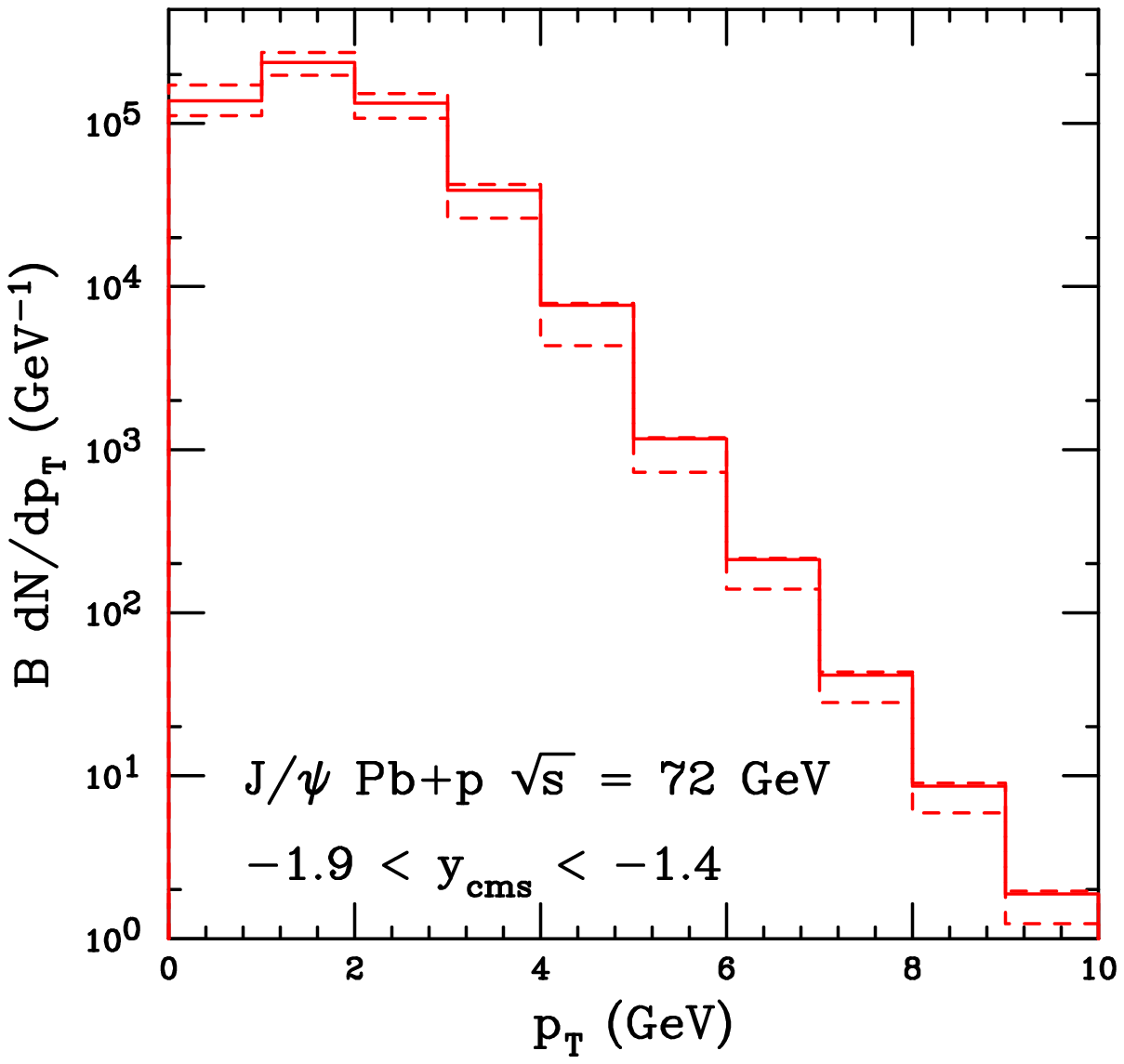} 
\end{center}
\caption{The predicted $J/\psi$ shadowing ratios (left) and rates (right)
as a function of center of mass rapidity (top) and $p_T$ (bottom) 
for Pb+$p$ collisions at $\sqrt{s_{_{NN}}} = 72$ GeV.
The solid curve in each plot is the central EPS09 NLO result
while the dotted curves outline the shadowing uncertainty band.
}
\label{Jpsi_Pbp}
\end{figure*}

\begin{figure*}[thp!]
\begin{center}
\includegraphics[width=0.495\textwidth]{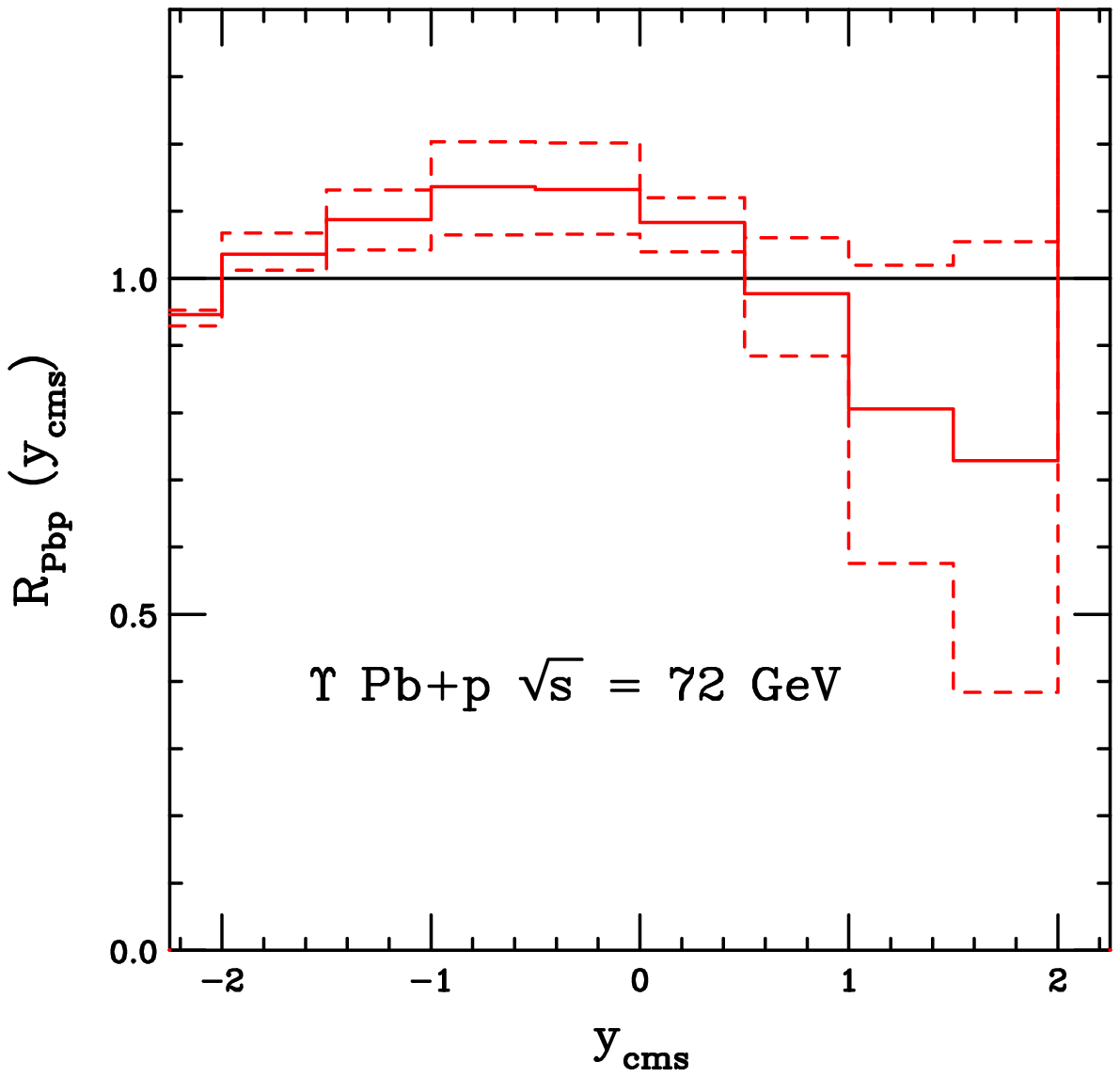}
\includegraphics[width=0.495\textwidth]{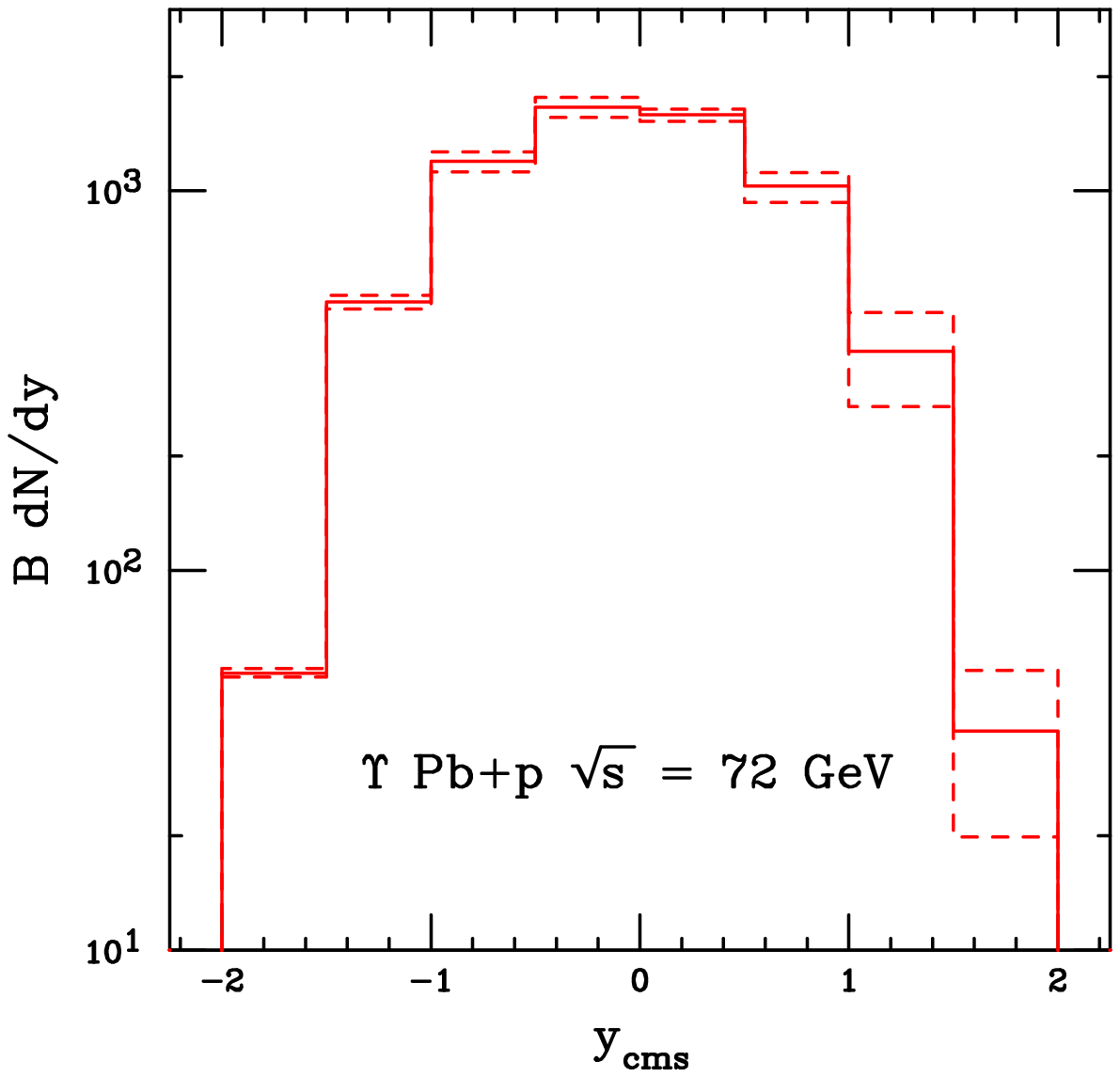} 
\includegraphics[width=0.495\textwidth]{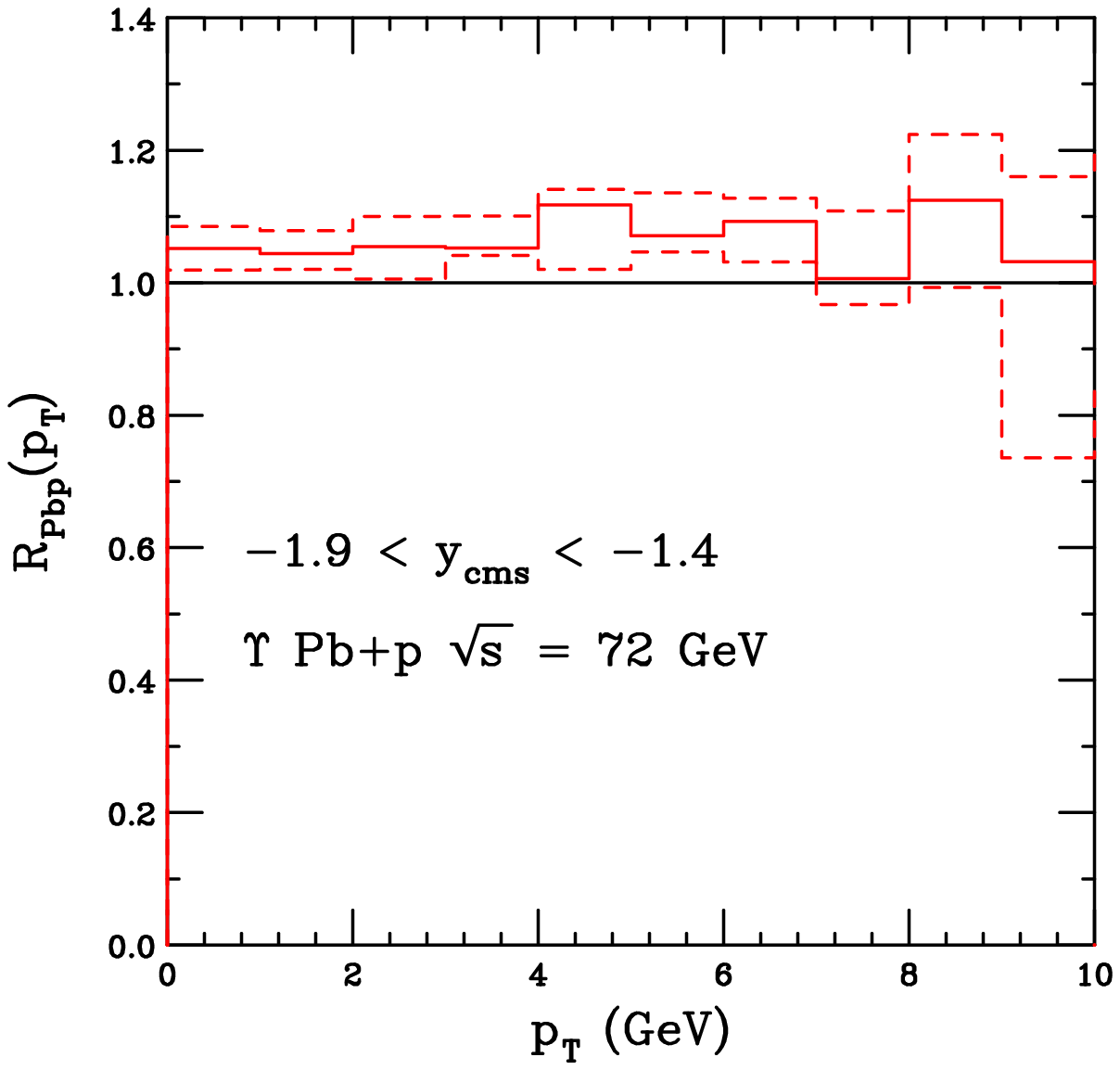}
\includegraphics[width=0.495\textwidth]{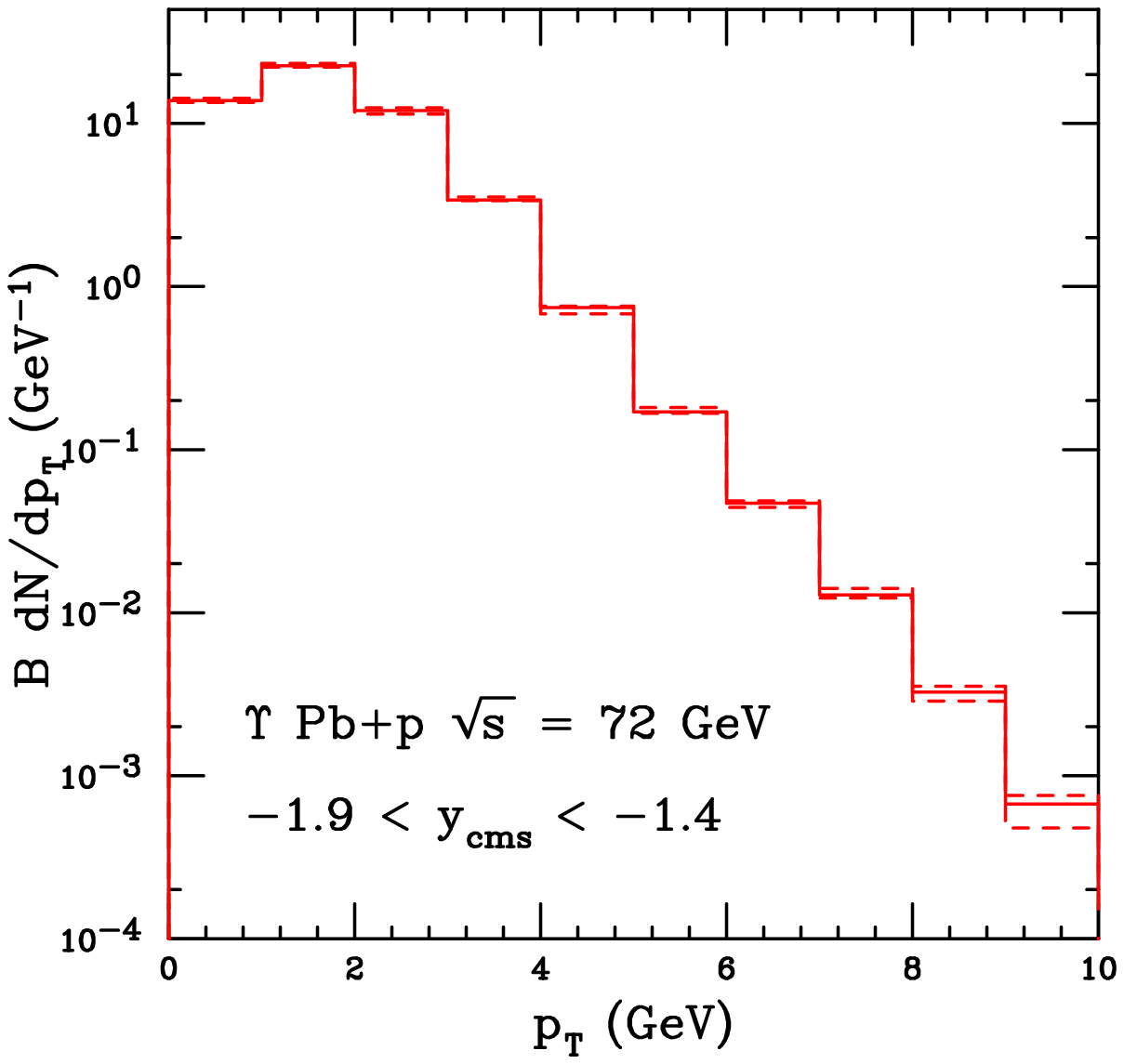} 
\end{center}
\caption{The predicted $\Upsilon$(1S) shadowing ratios (left) and rates (right)
as a function of center of mass rapidity (top) and $p_T$ (bottom)
for Pb+$p$ collisions at $\sqrt{s_{_{NN}}} = 72$ GeV.
The solid curve in each plot is the central EPS09 NLO result
while the dotted curves outline the shadowing uncertainty band.}
\label{Ups_Pbp}
\end{figure*}

In this section, the results for $J/\psi$ and $\Upsilon$ shadowing in Pb$+p$
collisions at $\sqrt{s_{_{NN}}} = 72$ GeV are presented.  Figure~\ref{Jpsi_Pbp}
shows the results for $J/\psi$ while Fig.~\ref{Ups_Pbp} shows the $\Upsilon$
results.  In both cases, the left-hand side shows the ratios 
$R_{{\rm Pb}p}$ as a function of $y_{\rm cms}$ (top) and $p_T$ in the 
$-1.9 < y_{\rm cms} < -1.4$ rapidity bin (bottom).  The rates
to dileptons in the rapidity acceptance, assuming a 1 m long
liquid hydrogen target, 
are shown on the right-hand side of the figures and in the bottom
parts of Tables~\ref{RateTable} and \ref{RateTable_pT}.  
The lower cross sections at
this reduced energy still result in rather high rates for the long liquid
hydrogen target, at least at midrapidity.

In these kinematics, the rapidity bin $-1.9 < y_{\rm cms} < -1.4$ now
corresponds to the more typical fixed-target kinematics with the lead nucleus
at lower $x_1$.  Here the ratio $R_{{\rm Pb}p}(y_{\rm cms})$ is reversed.  The $x$
range for the $J/\psi$ is in the higher $x$ end of the shadowing region
while the $\Upsilon$ is just entering the antishadowing region, recall
Fig.~\ref{EPS09fig}.

The antishadowing peak for $J/\psi$ in the upper left side of 
Fig.~\ref{Jpsi_Pbp} is actually just at forward rapidity instead of in the
$y_{\rm cms} < 0$ region.   (The full center-of-mass 
rapidity range for $J/\psi$ production at this energy is $|y_{\rm cms}| < 3.3$.)
Within the chosen rapidity bin, 
the $p_T$-dependent ratio 
has the largest uncertainty at low $p_T$ where there is still some shadowing.
However, at $p_T > 5$ GeV, the $x$ values move somewhat into the antishadowing
region (lower left panel). 

The $J/\psi$ rates for this system are still high,
see the lower half of Table~\ref{RateTable}.  Thanks to the length of the 
H$_2$ target, for the lead beam the $p_T$-integrated
rates in this configuration are 
still on the order of $10^6$ in the AFTER@LHC acceptance.  
The $p_T$-dependent rates show that the statistics become poor for the $J/\psi$
at $p_T > 7$ GeV.  The rates at this energy are helped somewhat since there is 
antishadowing for $p_T > 8$ GeV while there is strong
shadowing at $\sqrt{s_{_{NN}}} = 115$ GeV, see Fig.~\ref{Jpsi_pPb}.

The situation with $\Upsilon$(1S) is similar, see Fig.~\ref{Ups_Pbp} and
Table~\ref{RateTable}.  (The rapidity range for $\Upsilon$ production is
$|y_{\rm cms}| < 2.4$ so that again the AFTER@LHC acceptance is on the edge
of the $\Upsilon$ range.)  The shadowing (or antishadowing) effect 
is on the order of a few percent.  While there are a few thousand
$\Upsilon$ in a year at midrapidity, the rate in the AFTER@LHC acceptance is
rather low, under 100 per year, as shown in the lower part of the figure
and in Table~\ref{RateTable_pT}.  Indeed, there is effectively no rate for
the $\Upsilon$ rate for $p_T > 4$ GeV.

\section{Conclusions}
\label{summary}

We have only presented a bare minimum of the rates for the breadth of
quarkonium studies possible at AFTER@LHC.  The fixed-target configuration,
especially for long runs with the dedicated proton beam, allows 
detailed measurements with a range of nuclear targets.  We have only
shown the Pb results here because the larger nuclear mass number produces
what is expected to be the maximum effect due to shadowing.

The large $x$
region available for nuclear targets in the AFTER@LHC kinematics with a proton
beam has the unique
capability to make unprecedented studies of this heretofore unexplored range.
The AFTER@LHC measurements would bridge the gap between the 
dedicated fixed-target
experiments in the range $17.2 \leq \sqrt{s_{_{NN}}} \leq 41$ GeV and the
d+Au and upcoming $p+A$ collider experiments at RHIC, albeit in an
$x$ range never before studied.  

In the AFTER@LHC configuration with a Pb beam, the rates are 
smaller, though still significant, and the more conventional 
$x$ range is probed.  


\section*{Acknowledgments}
This work was performed under the auspices of the U.S.\
Department of Energy by Lawrence Livermore National Laboratory under
Contract DE-AC52-07NA27344.  I also thank the Institute for Nuclear Theory 
at the University of Washington, where this work was initiated, for 
hospitality.

\end{document}